\documentclass{aa}  
\usepackage{xcolor}
\newcommand{\tildea}[1]{\overset{\sim}{#1}}

\newcommand{\rev}[1]{#1}

\usepackage{graphicx}
\usepackage{mathtools} 
\usepackage{txfonts}

\begin{document}

   \title{The survivorship bias of protoplanetary disc populations}
   \subtitle{Internal photoevaporation causes an apparent increase of the median total disc mass with time}

   \author{Lorenzo A. Malanga
          \inst{1,2}\thanks{\email lorenzo.malanga@unimi.it}
          \and
          Giovanni P. Rosotti\inst{1}
          \and
          Giuseppe Lodato\inst{1}
          \and
          Alice Somigliana\inst{2,3,4}
          \and
          Carlo F. Manara\inst{2}
          \and
          Claudia Toci\inst{2}
          \and Leonardo Testi\inst{5,6}
          }

   \institute{Dipartimento di Fisica 'Aldo Pontremoli', Università degli Studi di Milano, Via Celoria 16, I-20133 Milano, Italy
             \and
             European Southern Observatory, Karl-Schwarzschild-Strasse 2, D-85748 Garching bei München, Germany
             \and Max-Planck Institute for Astronomy (MPIA), Königstuhl 17, 69117 Heidelberg, Germany
             \and Fakultat für Physik, Ludwig-Maximilians-Universität München, Scheinersts. 1, 81679 München, Germany
             \and Dipartimento di Fisica e Astronomia, Universita’ di Bologna, Via Gobetti 93/2, I-40122 Bologna, Italy.
             \and 
              INAF-Osservatorio Astrofisico di Arcetri, Largo E. Fermi 5, I-50125 Firenze, Italy.\\
             }

   \date{Received March 26, 2025; accepted May 7, 2025}

   \abstract{The evolution of protoplanetary discs has a substantial impact on theories of planet formation. 
   To date, \rev{neither} of the two main competing evolutionary models, namely the viscous-photoevaporative paradigm and the MHD winds model, has been ruled out by observations.
   Due to the high number of sources observed by large surveys, population synthesis is a powerful tool to distinguish the evolution mechanism in observations.
   We explore the evolution of the mass distribution of synthetic populations under the assumptions of turbulence-driven accretion and dispersal caused by internal photoevaporation.
   We find that the rapid removal of light discs often results in an apparent increase of the median mass of the survived disc population.
   This occurs both when the disc properties are independent of each other, and when typical correlations between these quantities and the stellar mass are assumed.
   Furthermore, as MHD wind-driven accretion rarely manifests the same feature, this serves as a signature of the viscous-photoevaporative evolution when dispersal proceeds from inside-out.
   Therefore, we propose the evolution of the median mass as a new method to distinguish this model in observed populations.
   This survivorship bias is not shown by the median accretion rate, which, instead, decreases with time.
   Moreover, we introduce a new criterion that estimates the disc lifetime as a function of initial conditions and an analytical relation to predict whether internal photoevaporation triggers an inside-out or an outside-in dispersal. 
   We verify both analytical relations with numerical simulations.
   
   }

\titlerunning{The survivorship bias of protoplanetary disc populations}
\authorrunning{Malanga, L. et al.}
\maketitle

\section{Introduction}
\label{intro}
Protoplanetary discs are the cradles of planets and their evolution and dispersal strongly affects the process of planet formation; therefore, studying the mechanism that drives accretion in discs is fundamental to understand the formation of planetary systems.
Traditionally, accretion is explained by the redistribution of angular momentum throughout the disc \citep{LyndenBellPringle1974, Pringle1981}. The transport of angular momentum is triggered by the presence of turbulence, parameterized by the $\alpha$ prescription of \cite{ShakuraSunyaev1973}.
In this scenario, disc dispersal is explained by photoevaporative winds launched from the disc surface when heated by stellar energetic radiation \citep{Hollenbach1994, Alexander2014, Ercolano2017}.
Numerical simulations have shown that X-ray and Far Ultraviolet (FUV) photons coming from the central star can trigger a thermal wind, \rev{whose emission peaks at 1-10 au} \citep{Gorti2009, Owen2010, Nakatani2018, Picogna2019, Sellek2024b}.
Moreover, models that combine viscous accretion and dispersal due to internal photoevaporation \citep{Clarke2001, Alexander2006,Owen2012} can reproduce the observed disc properties and disc dispersal time \citep{Hernandez2007, Fedele2010}.
However, recent observational constraints pointing to low turbulence in discs (see \citealt{Rosotti2023} for a review) represent a challenge for the viscous paradigm.
The best alternative to the classic viscous framework is the magneto-hydrodynamical (MHD) winds model, proposed by \cite{BlandfordPayne1982} and refined by recent studies (see \citealt{Lesur2021} for a review).
This theory suggests that angular momentum and mass are both removed by MHD winds.

The advent of a new generation of instruments, such as the Atacama Large Millimeter and submillimeter Array (ALMA) and the Very Large Telescope (VLT) has allowed to perform numerous surveys of different star-forming regions (see \citealt{Manara2023} for a review).
These surveys have measured various disc macroscopical quantities, such as the dust mass (e.g., \citealt{Miotello2016,Barenfeld2016, Ansdell2016, Ansdell2017, Bergin2017, Cazzoletti2019, Testi2022}), the mass accretion rate (e.g., \citealt{Alcala2014, Alcala2017, Manara_2017, Almendros-Abad2024}) and the disc radius, \rev{both with dust (e.g., \citealt{Tazzari2017,Andrews2018, Hendler2020, Sanchis2020}) and gas (e.g., \citealt{Ansdell2018, Najita2018, Sanchis2020}) measurements. The millimetric flux, which traces the dust mass, is often used as a proxy to measure disc gas mass, although the conversion presents multiple challenges (see \citealt{Miotello2022} for a review).} 

The statistical relevance of these samples represents a unique opportunity to test our evolutionary models with the tool of population synthesis.
However, \rev{neither} of the two frameworks has been ruled out by the observations yet.
Indeed, distinguishing the signatures of the two models in the observations has proven to be challenging.
For instance, viscous models predict ``viscous spreading'', the increase in radius of gas discs due to the transport of angular momentum, although the relationship between observed gas radius and the expanding disc radius is not trivial \citep{Toci2023}.
Since MHD winds simulations do not show this feature (e.g., \citealt{Zagaria2022}, who modeled the effect of MHD winds on dust discs), detecting an increasing radial size of the discs would be a signature of viscous evolution.
However, to detect it we would need a higher sensitivity and a larger sample size \citep{Trapman2020}.
Moreover, detecting it using dust as a proxy is still complicated, mainly due to the presence of radial drift and substructures \citep{Rosotti2019, Toci2021, Delussu2024}.

\cite{Lodato2017} showed that the purely viscous model inherently reproduces the observed accretion rate ($\dot{M}$) - disc mass ($M$) correlation (see also \citealt{Mulders2017}), while \cite{Somigliana2022} extensively studied how a correlation between the initial disc parameters and the stellar mass $M_\star$ can affect the evolution of the slopes.
\cite{Somigliana2020} introduced internal photoevaporation to the population synthesis model of \cite{Lodato2017} and showed for the first time that the removal of light discs causes older population to appear more massive.
On the MHD side, \cite{Tabone2022} showed that the properties of observed populations and the observed disc fraction can also be reproduced assuming a purely wind-driven evolution.
In addition, \cite{Zallio2024} managed to reproduce the observed $\dot{M} - M$ correlation with the MHD winds model, assuming a power law correlation between the initial mass and the accretion timescale.

Various population synthesis models have attempted to develop a method to disentangle the two models in observations.
\cite{Somigliana2024} illustrated how MHD evolution causes a different evolution of the slopes of the $M - M_\star$ and $\dot{M}-M_\star$ correlations with respect to the viscous-photoevaporative scenario. However, they emphasized that the current sample size does not allow us to distinguish between the two evolutionary models.
A similar argument is made by \cite{Alexander2023}, who state that a larger sample size, with $N>300$ would enable us to distinguish the two models from the $\dot{M}$ distribution.
\cite{Somigliana2023} investigated the spread of the $\dot{M}/M$ ratio as a proxy to disentangle between the two models from the observations. They found a better agreement with the MHD winds model, although the observational spread does not allow us to rule out the viscous scenario.
The picture is further complicated by the presence of external photoevaporation \citep{Coleman2024}, whose effect on the disc masses and radii is not negligible even for moderately irradiated environments \citep{Anania2024}.

All of these endeavors highlight that measuring the disc gas masses is crucial to understand disc evolution from observed populations.
For this scope, the new large program AGE-PRO (Zhang et al. in prep.).
have measured disc masses for three star-forming regions of different ages, using CO and N$_2$H$^+$ as a proxy \citep{Trapman2022}. 
Tabone et al. in prep. attempted to reproduce the observed population with both the MHD winds model and the viscous model and found that the latter fails to simultaneously reproduce both masses and accretion rates.
This result strengthens the support for the MHD winds model.
Data from the DECO (Cleeves et al. in prep.) large program will enhance the current sample size and allow us to verify this result and probe disc evolution. Hence, it is essential to identify the signatures of evolutionary models in disc populations and develop new proxies to distinguish them in observations.

In this paper, we explore the evolution of the mass distribution of disc populations when dispersal is considered, which has never been previously examined in the literature.
In particular, we investigate its behaviour under the assumptions of viscous evolution and dispersal driven by internal photoevaporation.
We obtain that the evolution of the median mass is a signature of this model and we propose it as a new proxy to distinguish between the viscous-photoevaporative framework and the MHD winds model in observed populations.
We also introduce a new criterion to predict a disc's lifetime based on its initial conditions.
In Section \ref{model} we present our analytical derivations, while in Section \ref{population_synthesis} we support them with numerical simulations.
In Section \ref{discussion} we discuss the implications of our results and compare them with different evolutionary pathways.

\section{Analytical model}
\label{model}

\subsection{Overview: the viscous-photoevaporative model}

In this section we provide a brief overview of the viscous model coupled with dispersal due to internal photoevaporation.
Accretion is explained by the radial transport of angular momentum, due to the presence of turbulence inside the disc. 
The gas surface density evolves according to the classic viscous evolution equation
\begin{equation}
    \frac{\partial \Sigma}{\partial t} = \frac{3}{R}\frac{\partial}{\partial R}\left(R^{1/2}\frac{\partial}{\partial R}(\alpha c_\text{s}H\Sigma R^{1/2})\right) - \dot{\Sigma}_\text{w}(R)\ ,
    \label{eqn:master_pe}
\end{equation}
where $R$ is the cylindrical radius and $\dot{\Sigma}_\text{w}$ the photoevaporative wind mass-loss rate per unit area.
The parameterization of viscosity $\nu$ follows the \cite{ShakuraSunyaev1973} prescription $\nu = \alpha c_\text{s} H$, where $c_\text{s}$ is the speed of sound, $H$ the scale height of the disc, and $\alpha$ a dimensionless parameter that quantifies turbulence, typically assumed constant throughout the disc. 
In the purely viscous scenario $\dot{\Sigma}_\text{w} = 0$ at all radii and equation \eqref{eqn:master_pe} admits an analytical ``self-similar'' solution \citep{LyndenBellPringle1974},
\begin{equation}
    \Sigma_\text{SS}(R,t) = \frac{M_0}{2 \pi R R_0}\left(1+\frac{t}{t_\nu}\right)^{-3/2}\exp\left({-\frac{R}{R_0 (1+t/t_\nu)}}\right) \ ,
    \label{eqn:self_similar}
\end{equation}
where $M_0$ is the initial disc mass, $R_0$ is the initial truncation radius and $t_\nu$ the viscous timescale, defined as $t_\nu =  {{R_0}^2}/3 \nu_c$, where $\nu_c = \nu(R = R_0)$.
Equation \eqref{eqn:self_similar} is valid under the assumption that viscosity scales as a power law as a function of radius,
\begin{equation}
    \nu = \nu_0 (R/R_0)^\gamma \ ,
    \label{eqn:viscosity}
\end{equation}
with $\gamma =1$ following the relation $T \propto R^{-1/2}$ \citep{Hartmann1998}, although solutions for different values of $0<\gamma<2$ also exist. We always make this assumption throughout the paper.
The total disc mass $M$ and the mass accretion rate onto the star $\dot{M}$ evolve with time, as:
\begin{equation}
    M(t) = M_0 \left(1+\frac{t}{t_\nu}\right)^{-1/2} \ ,
    \label{eqn:viscous_mass}
\end{equation}
\begin{equation}
    \dot{M}(t) = -\frac{\text{d} M}{\text{d} t} = \frac{M_0}{2 t_\nu}\left(1+\frac{t}{t_\nu}\right)^{-3/2} \ .
    \label{eqn:viscous_acc_rate}
\end{equation}
Equations \eqref{eqn:viscous_mass} and \eqref{eqn:viscous_acc_rate} highlight that both the disc mass and the accretion rate onto the star decrease with time due to viscous accretion.

\begin{figure}[ht!]
        \centering
        \includegraphics[
		width = \hsize]{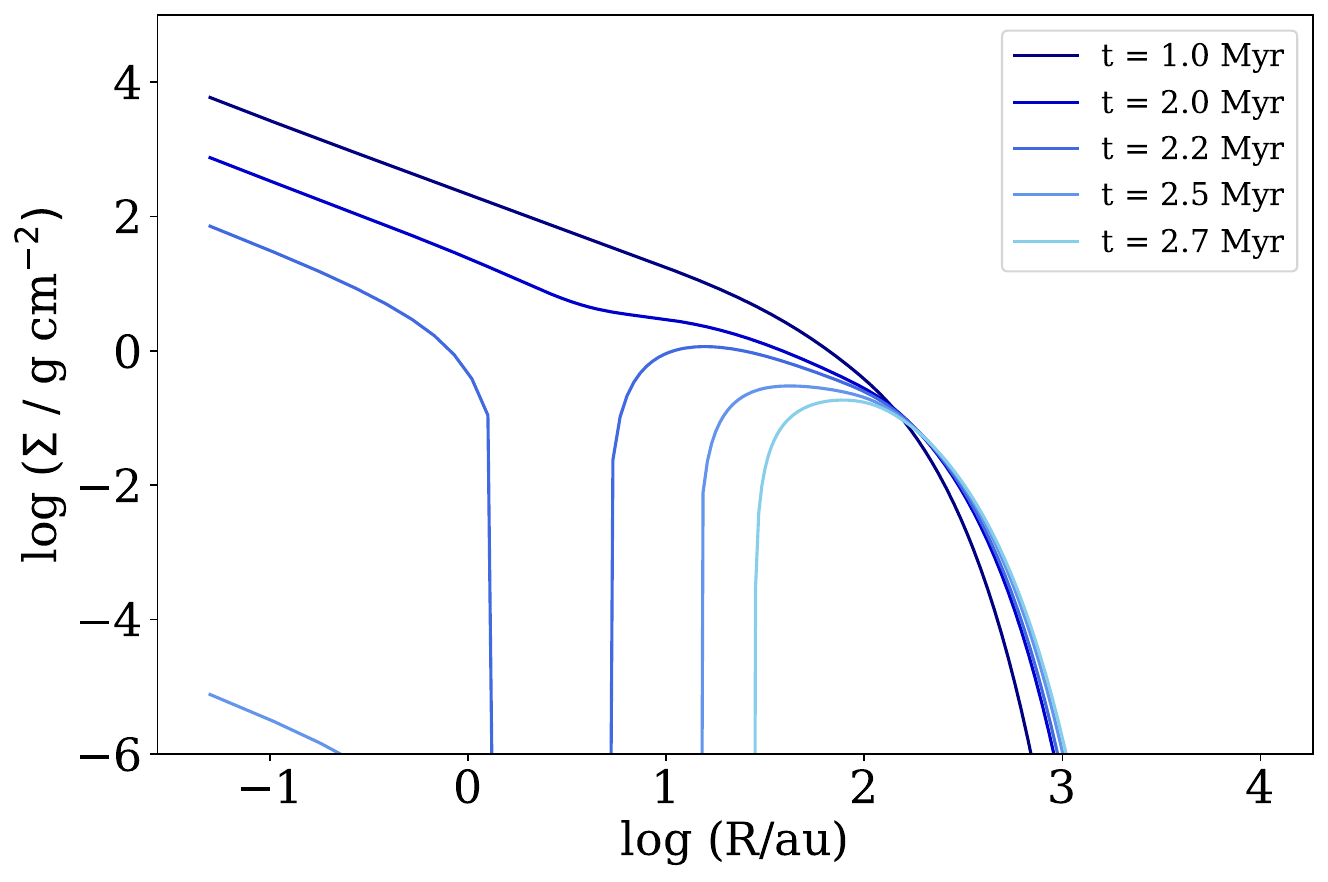}
        \caption{\small{Evolution of the surface density of a disc with time, obtained integrating equation \eqref{eqn:master_pe} with \texttt{Diskpop}. This disc has $M_0 = 1.5 \cdot 10^{-2}$ M$_\odot$, $R_0 = 32.8$ au, $\alpha = 10^{-3}$ and $\dot{M}_\text{w} = 2.5 \cdot 10^{-9}$ M$_\odot$ yr$^{-1}$. After the opening of the gap, the inner disc is dispersed within a short timescale.
        }}
        \label{fig:surfacedensity_gap}
    \end{figure}

\begin{figure}[ht!]
        \centering
        \includegraphics[
		width = \hsize]{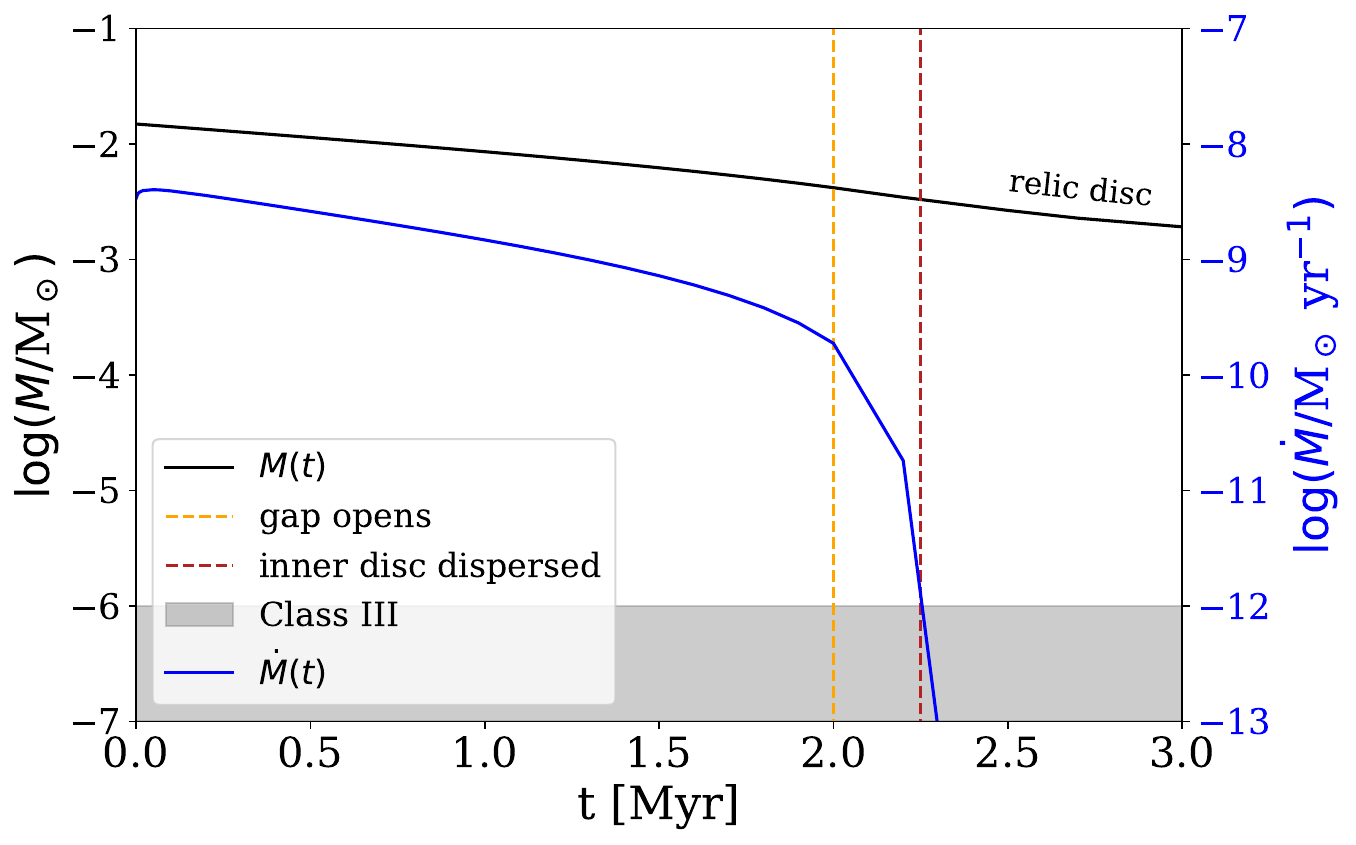}
        \caption{\small{Evolution of the accretion rate (blue) and mass (black) of a disc with time. The vertical yellow line indicates the instant when the gap opens, which triggers a steep decrease of $\dot{M}$. The red line corresponds to the transition between accreting and non-accreting. The grey region corresponds to lower values of disc mass and accretion rate than the respective observational threshold.
        }}
        \label{fig:mass_accrate}
    \end{figure}

Internal photoevaporation introduces a mass-loss term  
$\dot{\Sigma}_\text{w}(R)$ that can be computed with hydro-dynamic simulations \citep{Font2004, Alexander2006, Owen2012, Picogna2019} and it is often assumed to be solely dependent on $R$ and constant with time. 
Moreover, for X-ray driven photoevaporation, the radial profile of $\dot{\Sigma}_\text{w}(R)$ does not depend on the structure of the disc, as discussed in the literature \citep{Owen2010, Owen2012}.
The total wind mass-loss rate $\dot{M}_\text{w}$ is obtained by integrating $\dot{\Sigma}_\text{w}(R)$ along the radial component.
For simplicity, in this paper, we consider only the \cite{Owen2012} $\dot{\Sigma}_\text{w}$ numerical profile.
Numerical solutions of equation \eqref{eqn:master_pe} show that winds carve a cavity in the disc at around 5 au, corresponding to the peak of the mass-loss rate profile \citep{Picogna2021}. Figure \ref{fig:surfacedensity_gap} shows \rev{this} feature for a disc numerically evolved according to equation \eqref{eqn:master_pe} with the code \texttt{Diskpop} \citep{Somigliana2024}.
After the opening of the gap, the inner disc rapidly accretes onto the star, with a reduced viscous timescale \citep{Clarke2001};
\begin{equation}
    t_{\nu,\text{in}} = t_\nu \left(\frac{R_\text{gap}}{R_0}\right) \ ,
\end{equation}
where $R_\text{gap}$ is the radial location of the gap and $R_\text{gap} \ll R_0$ leads to $t_{\nu,\text{in}} \ll t_\nu$.
The drainage of the inner disc results in a steep decrease of the accretion rate, whereas the disc mass remains relatively constant (see Figure \ref{fig:mass_accrate}), due to the difficulty in dispersing the outer disc. The so-called ``relic disc'' problem is a well-known challenge of models that account only for internal photoevaporation as a source of disc dispersal without considering the direct irradiation of the disc after the gap opening \citep{Alexander2006, Owen2019, Robinson2025}.
Figure \ref{fig:mass_accrate} shows the evolution of the mass and the accretion rate with time of the same disc as Figure \ref{fig:surfacedensity_gap}.
Throughout this paper, we will refer to the dispersal time as the moment when either the disc mass or the accretion rate drop below the respective observational threshold, which we will define in Section \ref{population_synthesis}, and the object transitions to Class III. 

\subsection{The disc lifetime: state of the art}
\label{dispersal_time}

The fast dispersal of the inner disc results in an intrinsic two-timescale nature of the viscous-photoevaporative model, namely the ``ultraviolet switch'', introduced by \cite{Clarke2001}.
The first timescale corresponds to a slow viscous evolution, while the second one to a rapid clearance of the inner disc.
Due to the two mechanisms of gas depletion, the evolution of the disc mass can be described by the following relation until the gap opening:
\begin{equation}
    \frac{\text{d}M(t)}{\text{d}t} \sim -\dot{M}(t) - \dot{M}_\text{w} \ ,
\end{equation}
where the accretion rate $\dot{M}(t)$ is a decreasing function of time (see equation \ref{eqn:viscous_acc_rate}) while $\dot{M}_\text{w}$ is constant.
If we assume that $\dot{M}_0 \gg \dot{M}_\text{w}$, where $\dot{M}_0 \equiv \dot{M}(0)$, photoevaporation is negligible in the initial phase of the disc evolution, and equations \eqref{eqn:viscous_mass} and \eqref{eqn:viscous_acc_rate} adequately describe the evolution of the disc's macroscopical properties. 
However, as soon as $\dot{M}(t_\text{w}) = \dot{M}_\text{w}$, the effect of winds becomes relevant.
The time $t_\text{w}$, introduced by \cite{Clarke2001}, marks the transition between the accretion-dominated regime and the photoevaporation-dominated regime. As a consequence, due to the rapid photoevaporative removal of the inner disc, $t_\text{w}$ provides an estimate of the disc lifetime \citep{Clarke2001, Somigliana2020, Picogna2021}. The definition of $t_\text{w}$ implies 
\begin{equation}
    t_\text{w} = \frac{M_0}{2 \dot{M}_0}\left(\frac{\dot{M}_0}{\dot{M_\text{w}}}\right)^{\frac{2}{3}} = t_\nu \left(\frac{\dot{M}_0}{\dot{M_\text{w}}}\right)^{\frac{2}{3}} \gg t_\nu
    \label{eqn:tw}
\end{equation}
in the $t_\nu \ll t_\text{w}$ limit.

\subsection{The disc lifetime: a new criterion}
\label{dispersal_time_2}

Equation \eqref{eqn:tw} identifies the dispersal time as the moment when photoevaporation becomes the dominant driver of mass-loss on a global scale, a condition we will refer to as the ``global criterion'' in this work. 
However, we can easily see that this criterion is not general; for example, if $\dot{M}_0 < \dot{M}_\text{w}$, a scenario that cannot be excluded a priori, the criterion would predict a vanishingly small lifetime.
This is because equation \eqref{eqn:tw} does not take into account the finite time required for photoevaporation to deplete the surface density and open a gap, which, as illustrated by Figure \ref{fig:mass_accrate}, is the event that triggers the rapid dispersal of the inner disc. To derive this timescale, we need to make use of local quantities, such as the surface density $\Sigma$ and the local mass-loss rate $\dot{\Sigma}_\text{w}$.
Therefore, we introduce a ``local criterion'' that estimates the instant at which photoevaporative winds become locally relevant enough to cause the opening of a gap and the subsequent dispersal of the inner disc. 

We can write the evolution of $\Sigma (R,t)$ as a sum of a viscous term and a photoevaporative term:
\begin{equation}
    \Sigma (R,t) \sim \Sigma_\text{SS}(R,t) - \dot{\Sigma}_\text{w}(R)t \ ,
    \label{eqn:viscous_and_photoevaporative}
\end{equation}
where $\Sigma_\text{SS}(R,t)$ is given by equation \eqref{eqn:self_similar}.
We stress that equation \eqref{eqn:viscous_and_photoevaporative} is not a solution of equation \eqref{eqn:master_pe}, but rather an approximation.
We define the local dispersal timescale $t_\text{loc}$ as
\begin{equation}
    t_\text{loc} = \frac{\Sigma_\text{SS}(R_\text{gap}, t_\text{loc})}{\dot{\Sigma}_\text{w}(R_\text{gap})} \ ,
    \label{eqn:t_loc_sigma}
\end{equation}
which is the time at which equation \eqref{eqn:viscous_and_photoevaporative} vanishes at $R = R_\text{gap}$, which corresponds to the minimum of $\Sigma(R)/\dot{\Sigma}_\text{w}(R)$.
By considering the $\dot{\Sigma}_\text{w}(R)$ profile obtained by \cite{Owen2012} we find the local criterion of disc dispersal;
\begin{align}
    t_\text{loc} \sim  \ \rev{4.4} \ \text{Myr} & \left(\frac{M_0}{10^{-2}\ \text{M}}_\odot\right)^{3/5}\left(\frac{M_\star}{{\text{M}}_\odot}\right)^{3/5}\left(\frac{\dot{M}_0}{10^{-8} \ \text{M}_\odot \text{yr}^{-1}}\right)^{-1/5} \notag \\
    \times & \left(\frac{\dot{M}_\text{w}}{10^{-9}\ \text{M}_\odot \text{yr}^{-1}}\right)^{-2/5}\left(\frac{\alpha}{10^{-3}}\right)^{-2/5} \ ,
    \label{eqn:t_loc}
\end{align}
where 
\begin{equation}
    R_\text{gap} = 3.6 \ \text{au} \left(\frac{M_\star}{\text{M}_\odot}\right) 
    \label{eqn:peak_sigmadot}
\end{equation}
and 
\begin{equation}
    \dot{\Sigma}_\text{w}(R_\text{gap}) = \ 2.8 \cdot 10^{-11} \ \text{M}_\odot \ \text{yr}^{-1} \ \text{au}^{-2} \ \frac{1}{2 \pi}\left(\frac{M_\star}{\text{M}_\odot}\right)^{-2} \left(\frac{\dot{M}_\text{w}}{10^{-9} \ \text{M}_\odot \ \text{yr}^{-1}}\right) \ .
\end{equation}
We now have two criteria that estimate the disc lifetime based on the initial disc parameters, $M_0$ and $\dot{M}_0$ and on quantities related to the central star, $\dot{M}_\text{w}$ and $M_\star$.
The advantage of the newly introduced criterion is that it is valid also for discs with initial accretion rate lower than $\dot{M}_\text{w}$.
Nevertheless, if accretion is still the dominant driver of the mass loss at $t_\text{loc}$, viscous replenishment will prevent the gap opening. 
This implies that, while both the global and local dominance of photoevaporation are necessary conditions, they are not sufficient to trigger the gap formation. Instead, both criteria need to be simultaneously satisfied. As a consequence, the most accurate approximation of the disc lifetime corresponds to the maximum between the two timescales,
\begin{equation}
    t_\text{disp} = \text{max}(t_\text{w}, t_\text{loc}) \ .
    \label{eqn:t_disp}
\end{equation}
In Section \ref{population_synthesis} we will prove the validity of this criterion by showing that it provides an accurate estimate of the lifetime of synthetic discs.

Both equations \eqref{eqn:tw} and \eqref{eqn:t_loc} highlight a positive scaling between the dispersal time and the initial disc mass, which suggests that light discs are expected to be dispersed earlier than massive discs.

In the following section, we will show that this feature leads to the increase of the median mass of accreting discs with time. 

\subsection{The evolution of the disc mass distribution}

Having derived how long lived discs are, we now turn to the problem of quantifying the median mass of a disc population. For the purely viscous case, this has been done in \cite{Somigliana2022}. We briefly recap it, and then extend it to the photoevaporative case.

\subsubsection{Recap: purely viscous scenario}

We consider a population of discs whose initial mass and accretion rate follow a log-normal distribution. Defining $m = \log (M/\text{M}_\odot)$, $\dot{m}_0 = \log (\dot{M}_0/\text{M}_\odot \ \text{yr}^{-1})$ and $\tau = \log (t/\text{yr})$, the viscous equations \eqref{eqn:viscous_mass} and \eqref{eqn:viscous_acc_rate} at $t \gg t_\nu$ can be rewritten as follows:
\begin{equation}
    m = \frac{3}{2} m_0 - \frac{1}{2}\dot{m}_0 - \frac{1}{2}\log 2 -\frac{1}{2} \tau \ .
    \label{eqn:viscous_accretion_log}
\end{equation}
The distribution of masses at age $t$ is thus retrieved by integrating over all initial masses and accretion rates, under the assumption that equation \eqref{eqn:viscous_accretion_log} is satisfied,
\begin{equation}
    \frac{\partial N}{\partial m} = \int \int \frac{\partial N}{\partial m_0} \frac{\partial N}{\partial \dot{m}_0} \delta (\dot{m}_0 - \dot{m}_{0,m}) \ \text{d}m_0 \ \text{d}\dot{m}_0 \ ,
    \label{eqn:mass_distrib_viscous}
\end{equation}
where $\partial N/\partial m_0$ and $\partial N/\partial \dot{m}_0$ are normal distributions and $\dot{m}_{0,m} = 3 m_0 -2m -\tau -\log 2$.
As demonstrated by \cite{Somigliana2022}\footnote{The slight differences between our derivation and the one by \cite{Somigliana2022} arise from the fact that we chose to highlight the initial accretion rate rather than the viscous time.}, the calculation of this integral leads to a further log-normal distribution centred in $\overline{m}$, which is obtained substituting the mean values of the $m_0$ and $\dot{m}_0$ distributions in equation \eqref{eqn:viscous_accretion_log}, and with $\sigma^2 = \frac{9}{4} {\sigma_{m_0}}^{2} + \frac{1}{4} {\sigma_{\dot{m}_0}}^{2}$.
As a consequence, under the assumption of a purely viscous evolution, an initial log-normal mass distribution keeps being log-normal for the entirety of the disc's lifetime, and its mean shifts to lower masses due to viscous mass depletion.

\subsubsection{Viscous-photoevaporative scenario}
\label{analytic_m_distribution}

We now extend the framework by including internal photoevaporation, which has not been previously explored in the literature.
Assuming that discs undergo a purely viscous evolution until the dispersal time $t_\text{disp}$, the terms inside the integral in equation \eqref{eqn:mass_distrib_viscous} remain valid even when internal photoevaporation is introduced.
However, there is a fundamental difference: both relations \eqref{eqn:tw} and \eqref{eqn:t_loc} imply a finite disc lifetime, i.e., the population loses discs over time.
We use them to obtain the minimum initial mass that is required for a disc not to be dispersed at age $t$. Therefore, we modify equations \eqref{eqn:tw} and \eqref{eqn:t_loc} so that $t_\text{disp} (M_0 = M_\text{0,MIN}) = t$. We obtain
\begin{equation}
    m_{0,\text{MIN}} = \tau + \log 2 + \frac{1}{3}\dot{m}_0 + \frac{2}{3}\dot{m}_\text{w}
    \label{eqn:m0_min_tw}
\end{equation}
for the global criterion and 
\begin{equation}
    m_{0,\text{MIN}} = \frac{5}{3}\tau - \frac{5}{3}\tau_0 + \frac{1}{3}\dot{m}_0 + \frac{2}{3}\dot{m}_\text{w} 
 + \frac{2}{3} \log \alpha
 - m_\star \ 
    \label{eqn:m0_min_tloc}
\end{equation}
for the local criterion, with $\tau_0 = 1.44$ (see equation \ref{eqn:t_loc}) \rev{and $\dot{m}_\text{w} \equiv \log(\dot{M}_\text{w}/ \text{M}_\odot \ \text{yr}^{-1})$}.
We obtain the mass distribution by setting $m_{0,\text{MIN}}$ as the lower boundary of the integral
\begin{equation}
    \frac{\partial N}{\partial m} = \int^{+\infty}_{m_{0,\text{MIN}}(\tau)} \text{d}m_0  \int \frac{\partial N}{\partial m_0} \frac{\partial N}{\partial \dot{m}_0} \delta (\dot{m}_0 - \dot{m}_{0,m}) \ \text{d}\dot{m}_0 \ .
    \label{eqn:mass_distrib_vpe}
\end{equation}
\rev{We derive and show the analytical relation in the Appendix}.

\begin{figure}[ht!]
        \centering
        \includegraphics[
		width = \hsize]{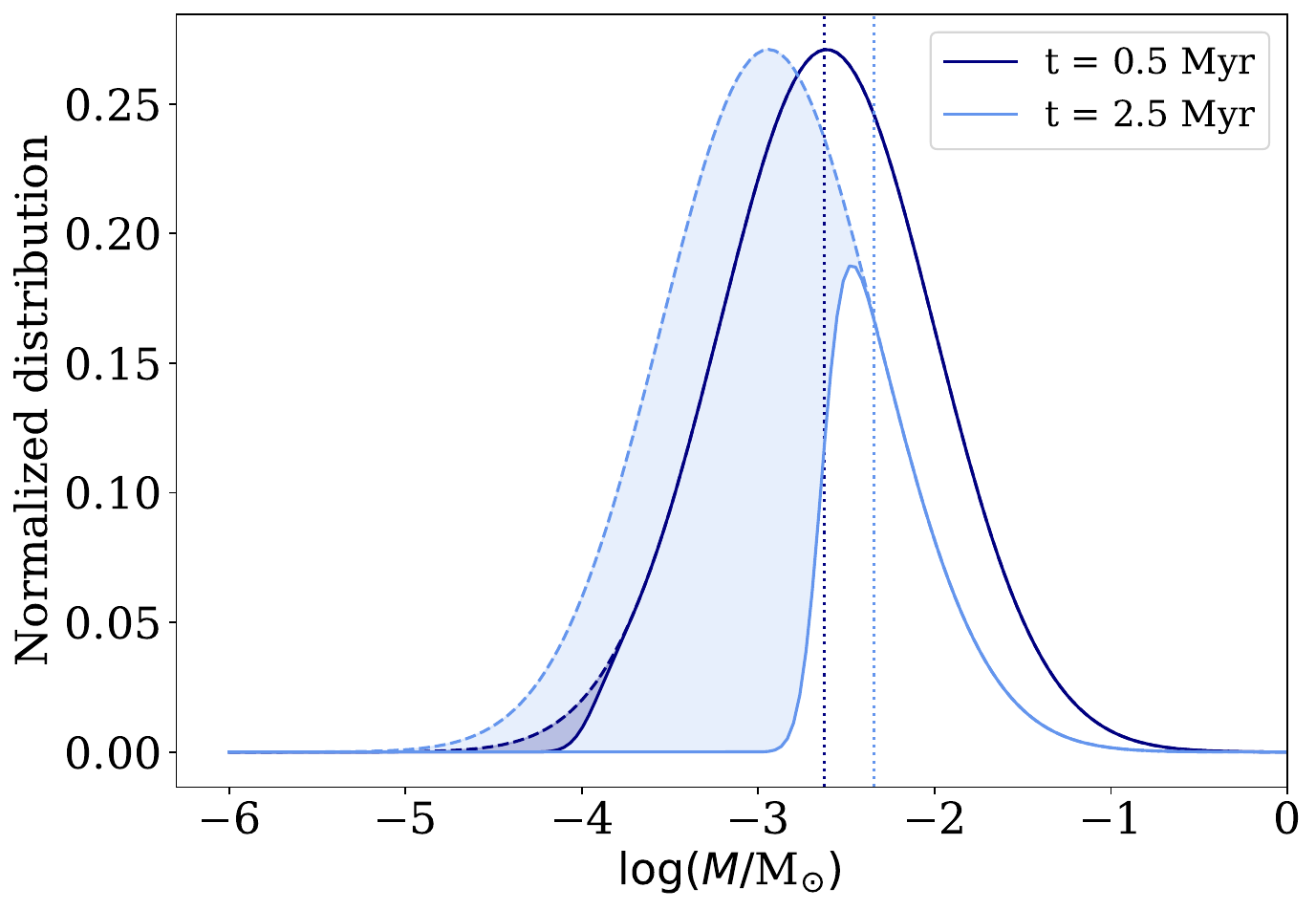}
        \caption{\small{Disc mass distribution at 0.5 Myr and 2.5 Myr with the viscous-photoevaporative model (solid), compared with purely viscous distributions (dashed).
        The dotted vertical lines represent the medians of the distributions of survived discs, whereas the hatched regions correspond to the dispersed discs. This population has $\langle \log( M_0 /$M$_\odot ) \rangle$ = $-2.5$, $\langle \log( \dot{M}_0 /$M$_\odot$ yr$^{-1} )\rangle $ = $-8.3$ and $\langle \log( \dot{M}_\text{w} /$M$_\odot$ yr$^{-1} )\rangle $ = $-9.6$, where all quantities follow log-normal distributions with $\sigma = 0.5$ dex.
        }}
        \label{fig:cutgaussian}
    \end{figure}

Based on the considerations made in Section \ref{dispersal_time_2}, we take as $m_{0,\text{MIN}}$ the maximum between equations \eqref{eqn:m0_min_tw} and \eqref{eqn:m0_min_tloc}.
The obtained distribution is represented in Figure \ref{fig:cutgaussian} at 0.5 Myr and 2.5 Myr.
As the population evolves, the distribution is cut at low masses (hatched regions), due to the photoevaporative dispersal of light discs, which causes the median to increase with time.
Such behaviour was first noticed by \cite{Somigliana2020}, who studied the effect of internal photoevaporation on the distribution of populations in the $\dot{M} - M$ plane. 
This feature occurs because the removal of less massive discs, which follows equations \eqref {eqn:m0_min_tw} and \eqref{eqn:m0_min_tloc}, has always a steeper dependence on the age than viscous accretion (see equation \ref{eqn:viscous_accretion_log}) whose effect is to move the whole distribution to lower masses.
Hence, the fast removal of less massive discs causes older populations to appear more massive, not because of a mass gain but because the only discs left are the most massive. This is thus a case of survivorship bias.

\subsection{The effect of a correlation between the initial disc properties and $M_\star$}
\label{model_correlations}

In Section \ref{analytic_m_distribution} we found that the viscous-photoevaporative scenario predicts the median of the disc mass to increase with the age of the population, regardless the choice of initial parameters. 
We derived such feature by considering the scaling of the dispersal time with the initial mass and assuming that the other quantities do not depend on $M_0$. 
Here we explore the impact of a correlation between the initial disc properties and the stellar mass on results obtained in the previous section.

Numerous surveys of star-forming regions (see \citealt{Manara2023} for a review) revealed a pronounced correlation between the disc mass, obtained from the sub-mm continuum emission of the dust component, the accretion rate and the stellar mass.
In particular, discs around massive stars tend to be more massive and higher accretors.
Such correlations may be induced by disc evolution or may be due to initial conditions.
\cite{Somigliana2022} explored the latter scenario by generating synthetic populations where $M_0$ and $\dot{M}_0$ are correlated to $M_\star$ by power law relations
\begin{equation}
    M_0  \propto {M_\star}^ {\lambda_{\mathrm{m},0}} \ ,  
    \label{eqn:corr_mass}
\end{equation}
\begin{equation}
    \dot{M}_0  \propto {M_\star} ^ {\lambda_{\text{acc},0}} \ .
    \label{eqn:corr_accrate}
\end{equation}
Furthermore, numerical simulations \citep{Picogna2021} highlight that discs surrounding massive stars experience a more intense wind's mass-loss, correlated to $M_\star$ through a power law relation,
\begin{equation}
    {\dot{M}_\text{w}} \propto {M_\star}^{\lambda_\text{w}} \ ,
    \label{eqn:corr_mw}
\end{equation}
where we did not label $\lambda_\text{w}$ with the subscript ``0'' because, in our model, $\dot{M}_\text{w}$ is constant throughout the disc's lifetime.
Observational and theoretical constraints result in positive exponents, meaning that discs around massive stars not only tend to be more massive themselves but also tend to experience higher mass-losses due to both accretion and internal photoevaporation.
This may imply that, although having a larger initial mass, they may be shorter lived due to a more vigorous wind.

To assess whether the increase of the median disc mass is still expected in the viscous-photoevaporative model if we include a scaling of this properties with $M_\star$, we can substitute the power law relations \eqref{eqn:corr_mass}, \eqref{eqn:corr_accrate} and \eqref{eqn:corr_mw} into the equations of the dispersal time \eqref{eqn:tw} and \eqref{eqn:t_loc} and obtain the scaling of $t_\text{disp}$ with $M_0$;
\begin{equation}
    t_\text{disp} \propto {M_0}^\lambda \ ,
\end{equation}
where
\begin{equation} \lambda  = \begin{dcases}
    \frac{3 \lambda_{\mathrm{m},0} - \lambda_{\text{acc},0} - 2 \lambda_\text{w}}{3 \lambda_{\mathrm{m},0}} & \mbox{for the global criterion}\\ \label{eqn:t_disp_corr1}
    \frac{3 +3  \lambda_{\mathrm{m},0} - \lambda_{\text{acc},0} - 2 \lambda_\text{w}}{5 \lambda_{\mathrm{m},0}} & \mbox{for the local criterion}\ . 
\end{dcases}
\end{equation}
Light discs are the first to be depleted only if $\lambda > 0$. 
However, as shown in Section \ref{analytic_m_distribution}, to allow the median mass to increase with time, $M_{0,\text{MIN}}$ must scale with a steeper power law of $t_\text{disp}$ than accretion.
This condition is met when $0<\lambda < 2$ (see equation \ref{eqn:viscous_accretion_log}).
Hence, the slopes of the initial correlations with $M_\star$ play a crucial role in determining whether the median mass of a population increases with time.
To investigate whether this trend can be reproduced, we now consider an example based on the plausible values of the slopes reported in the literature.
Specifically, following the constraints of \cite{Somigliana2022}, who studied the evolution of the slopes of the correlations in synthetic population of viscously accreting discs, we can assume that $\lambda_{\mathrm{m},0}$ and $\lambda_{\text{acc},0}$ are in the following ranges:
\begin{align}
    \lambda_{\mathrm{m},0} & \in [1.2,2.1] \ , \\
    \lambda_{\text{acc},0} & \in [0.7,1.5] \ .
\end{align}
Moreover, \cite{Somigliana2022} found that, in order to reproduce the observed evolution of the slopes of the $M-M_\star$ and $\dot{M} - M_\star$ correlations, $\lambda_{\mathrm{m},0}$ needs to be larger than $\lambda_{\text{acc},0}$.
For what concerns the $\dot{M}_\text{w}$ correlation, we can assume $\lambda_\text{w} \sim 1$, following \cite{Picogna2021}.
With these constraints we obtain that the condition $0 < \lambda < 2$, which is necessary for the increase of the median disc mass of survived discs, still holds for both dispersal criteria.

\vspace{0.25cm}

In these sections we highlighted how the assumption of a rapid inner disc clearing, caused by internal photoevaporation, results in the increase of the median mass of the population with time. Such feature is expected in the simplest scenario when $M_0$, $\dot{M}_0$ and $\dot{M}_\text{w}$ are not correlated to $M_\star$. Moreover the constraints on the slopes $\lambda_i$ provided by the literature suggest that the increase of the median mass is a property of the viscous-photoevaporative model even when the initial disc properties are correlated with stellar mass.
In the next section we support this statement with numerical simulations and discuss its implications.

\section{Numerical simulations with \texttt{Diskpop}}
\label{population_synthesis}

Analytical arguments presented in Section \ref{model} relied on two main assumptions: firstly, we assumed that photoevaporation does not affect the disc evolution until $t_\text{disp}$, when it instantly depletes the disc; moreover, we assumed that the disc mass and the accretion rate are distributed log-normally at age zero.
In this section we show that populations of discs, numerically evolved with the viscous-photoevaporative model, and which do not rely on our first assumption, still manifest an increasing median mass. 

We used the 1D code \texttt{Diskpop}, presented by \cite{Somigliana2024}, designed to generate synthetic populations of discs.
We generated an initial population of Young Stellar Objects (YSOs), each of which \rev{consists} of a star and a disc.
We extracted the stellar mass $M_\star$ from an Initial Mass Function (IMF), which, in \texttt{Diskpop}, can be selected among various ones, while the key initial parameters of the disc ($M_0$, $\dot{M}_0$ and $\dot{M}_\text{w}$) were extracted from log-normal distributions with means determined by power-law correlations with $M_\star$ (see equations \ref{eqn:corr_mass}, \ref{eqn:corr_accrate}, \ref{eqn:corr_mw}).
\texttt{Diskpop} allows to select normalizations of the power laws, i.e., the values of $M_0$, $\dot{M}_0$ and $\dot{M}_\text{w}$ for a disc surrounding a 0.3 M$_\odot$ star (which in our case are also the mean values of the distributions), the exponents $\lambda_i$ and the spreads $\sigma_i$.
The code solves numerically the evolution equation \eqref{eqn:master_pe} for each disc and returns the values of $\Sigma$, $M$ and $\dot{M}$ at different snapshots.
We imposed a threshold of $10^{-12}$ M$_\odot$ yr$^{-1}$ for the accretion rate, based on observational detectability \citep{Fedele2010} and a threshold for the disc mass of $10^{-6}$ M$_\odot$.
If either $\dot{M}$ or $M$ decreases below its respective threshold, the synthetic YSO is assumed to transition to Class III and is removed from the disc population.
We chose $\alpha = 10^{-3}$ for all discs in the population.

\subsection{Verifiying the analytical predictions}
\label{verifying analytical predictions}

We can use \texttt{Diskpop} to verify that the maximum between $t_\text{loc}$ (see equation \ref{eqn:tw}) and $t_\text{w}$ (see equation \ref{eqn:t_loc}) predicts the disc lifetime of synthetic discs
and that a population that evolves according to equation \eqref{eqn:master_pe} can be effectively described by the analytical distribution \eqref{eqn:mass_distrib_vpe}.
To achieve this, we need to implement the same initial conditions used to derive the analytical relation.
Consequently, we generate a population of 200 YSOs where we extract the stellar mass from a log-normal distribution centered in 0.3 M$_\odot$ and with $\sigma = 0.5$ dex. 
For each YSO, we obtain the disc properties $M_0$, $\dot{M}_0$ and $\dot{M}_\text{w}$ with the procedure described previously but choosing a null spread for all parameters. 
The result is a fictional population where all disc properties are perfectly correlated with $M_\star$.
We choose $\lambda_{\mathrm{m},0} = 2$, $\lambda_{\text{acc},0} = 1$ and $\lambda_\text{w} = 1$. 
The physical meaning of these correlation is that discs surrounding massive stars are more massive, higher accretors and experience a higher mass-loss due to internal photoevaporation.
We know from equations \eqref{eqn:t_disp_corr1} that this combination satisfies $0<\lambda<2$ for both $t_\text{w}$ and $t_\text{loc}$, resulting in massive discs having a longer lifetime.
We subsequently evolve the population until 15 Myr removing YSOs with lower accretion rates or masses than the imposed thresholds.

\subsubsection{Verifying the disc dispersal criterion}

\begin{figure}[ht!]
        \centering
        \includegraphics[
		width=0.8\hsize]{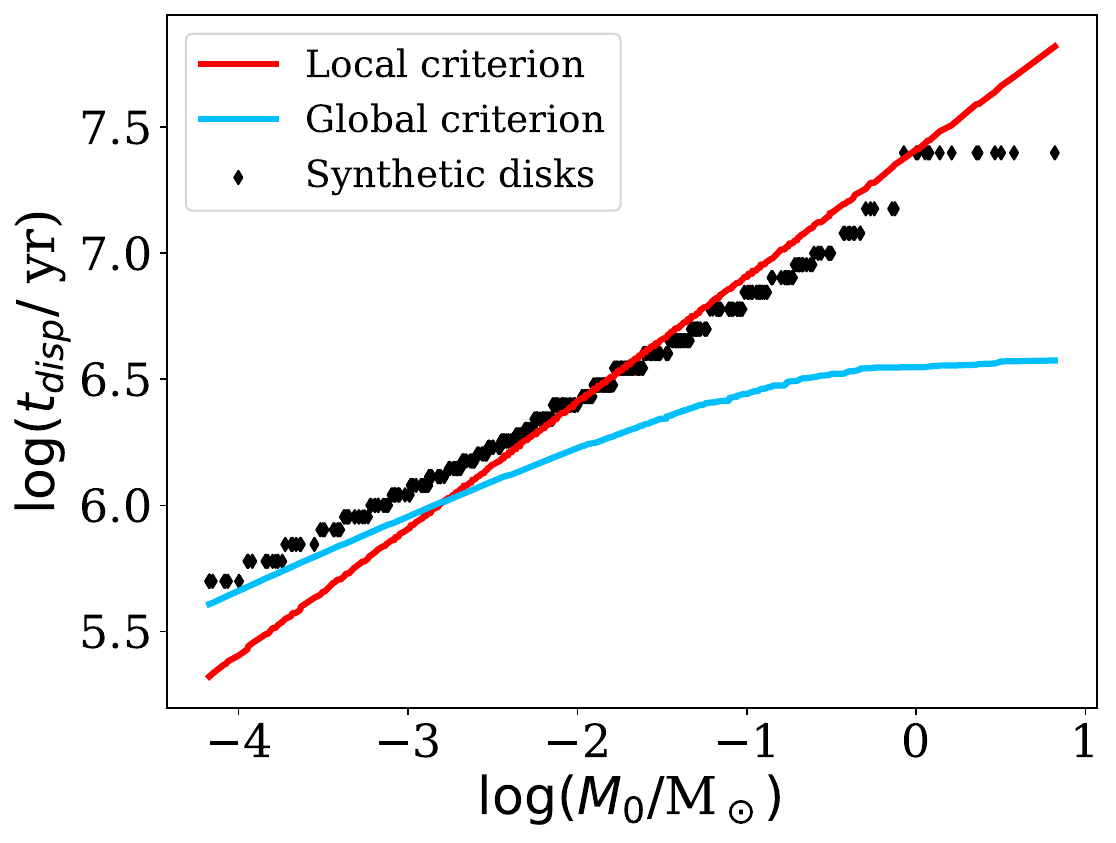}
        \caption{\small{Comparison between the synthetic disc lifetime as a function of $M_0$ and the analytical prediction.
        }}
        \label{fig:verification_tdisp}
    \end{figure}

We show in Figure \ref{fig:verification_tdisp} that the dispersal time of synthetic discs follows the analytical prediction.
The plot shows that massive discs follow the local criterion, while light discs follow the global criterion. The presence of a regime switch emphasizes the accuracy of our model. 
In this case, the lifetime of the majority of the discs is best approximated by the local criterion, although this can be reversed considering a different combination of initial conditions.

\subsubsection{Verifying the analytical mass distribution}

\begin{figure}[ht!]
        \centering
        \includegraphics[
		width=\hsize]{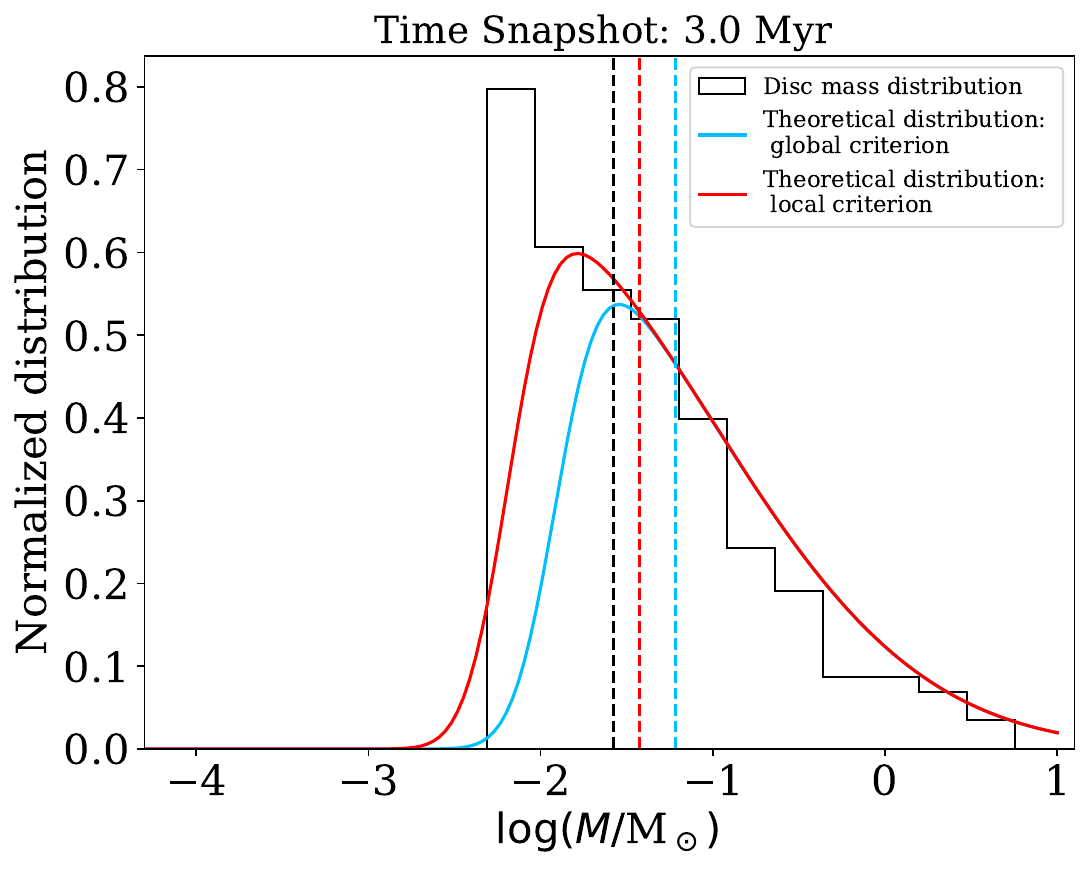}
        \caption{\small{Comparison between the synthetic mass distribution and the analytical prediction for both dispersal criteria. The vertical lines represent the medians of each distribution.
        }}
        \label{fig:mass_distributions}
    \end{figure}

\begin{figure}[ht!]
        \centering
        \includegraphics[
		width=\hsize]{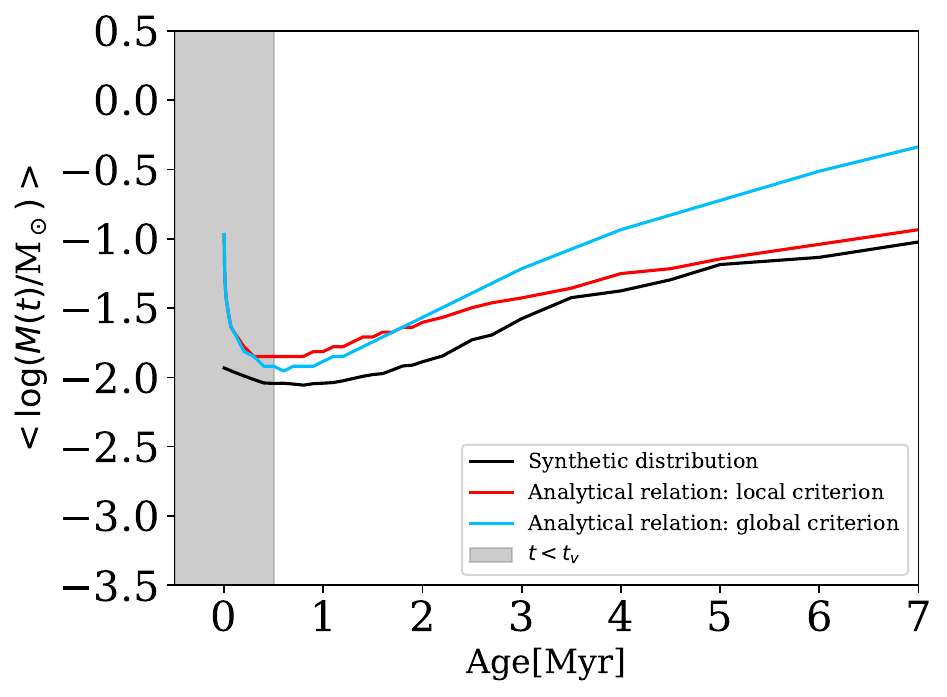}
        \caption{\small{Comparison between the evolution of the median of the synthetic mass distribution with time and the analytical prediction for both dispersal criteria. The shaded region corresponds to $t<t_\nu$, which is not considered in the model.
        }}
        \label{fig:medians}
    \end{figure}

Figure \ref{fig:mass_distributions} illustrates that the mass distribution of simulated discs aligns with the analytical prediction for both dispersal criteria. 
The distribution is strongly skewed since light discs are the first to be removed.
Furthermore, the median of survived discs in the synthetic population increases with time following the analytical expectations (see Figure \ref{fig:medians}). At early evolutionary stages ($t< 1$ Myr), the distribution is more accurately described by the global criterion; however, such relationship reverses in the later stages. This occurs because the lifetime of light discs is better approximated by the global criterion, while massive discs tend to follow the local criterion (see Figure \ref{fig:verification_tdisp}).
Figure \ref{fig:medians} also highlights a discrepancy between the model and the simulations at very early stages ($t \ll 1$ Myr), where the initial decrease of the median mass, due to viscous depletion, is less steep in the simulations than predicted by the analytical expectations. This discrepancy arises because we derived $m_{0,\text{MIN}}$ (see equations \ref{eqn:m0_min_tw} and \ref{eqn:m0_min_tloc}) under the assumption that $t \gg t_\nu$. Given that $t_\nu$ typically ranges from $10^5$ to $10^6 $ yr, the analytical distribution only describes evolved populations. 
We performed different simulation varying the slopes $\lambda_i$ and the normalizations of initial conditions and we confirmed that the increase of the median mass occurs only when predicted by the analytical model.

These simulations show that the evolution of the disc mass in populations evolved with the viscous-photoevaporative model is described by equation \eqref{eqn:mass_distrib_vpe}. 

\subsection{Generating a realistic disc population}
\label{realistic pop}

In the previous section we generated a population with ad-hoc initial conditions to verify the results obtained analytically.
In this section we exploit \texttt{Diskpop} in its intended purpose: generating realistic synthetic disc populations.
We show that the distinctive feature of the viscous-photoevaporative model that we presented in Section \ref{model} also occurs when more plausible initial distribution are implemented.
Firstly, \rev{for 200 discs}, we extract $M_\star$ from the Kroupa IMF \citep{Kroupa2001}, which, unlike the log-normal distribution used previously, favors lower stellar masses.
As for disc parameters distributions, we set $\lambda_{\mathrm{m},0} = 2$, $\lambda_{\text{acc},0} = 1$ and $\lambda_\text{w} = 1$ as slopes and a spread of 0.5 dex.
To ensure that our synthetic population matched the observed fraction of discs as a function of age \citep{Hernandez2007, Fedele2010}, which indicates an average disc lifetime of 2-3 Myr, we tailored the normalization of initial correlation so that $\rev{\langle t_\text{loc} \rangle } \sim$ 2.5 Myr.
\rev{Equation \eqref{eqn:t_loc} shows that there are infinite combinations that satisfy this condition.
Among all the possible combinations}, we chose \rev{the following set of parameters}: $\langle M_0 \rangle = 5 \cdot 10^{-3}$ M$_\odot$, $\langle \dot{M}_0 \rangle = 5 \cdot 10^{-9}$ M$_\odot$ yr$^{-1}$ and $\langle \dot{M}_\text{w} \rangle = 6 \cdot 10^{-10}$ M$_\odot$ yr$^{-1}$. \rev{This choice not only satisfies the constraint that we imposed on $t_\text{loc}$, but also allows us to qualitatively reproduce the observed disc masses and accretion rates.
However, we verified that any other combination would have equally served the purpose of this Section.}

\begin{figure}[ht!]
        \centering
        \includegraphics[
		width=\hsize]{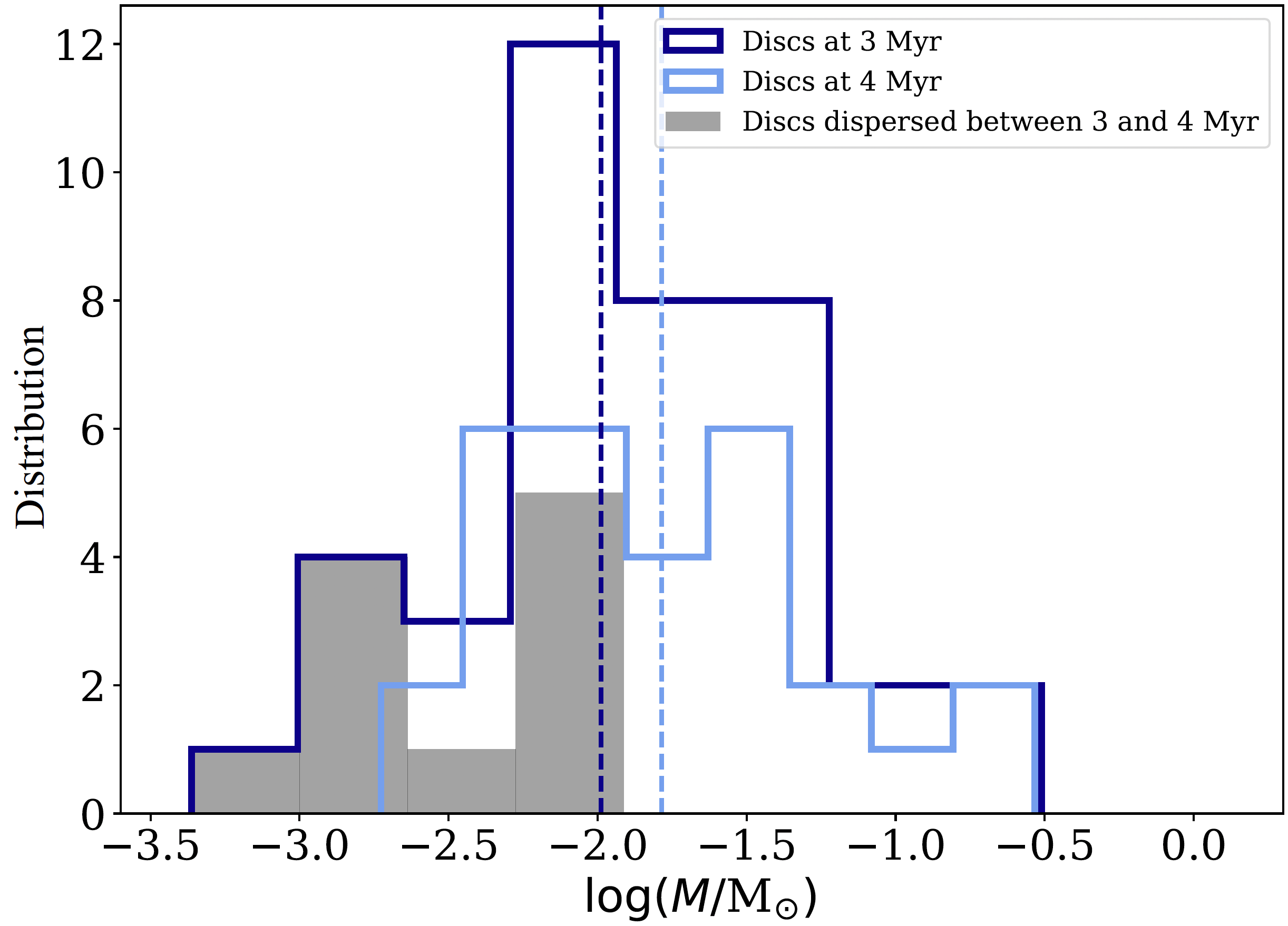}
        \caption{\small{Histogram of the disc mass distribution at 3 Myr (dark blue) and 4 Myr (light blue). The dashed vertical lines are the medians of the distributions. The shaded histogram represents the mass distribution of discs that are dispersed between 3 and 4 Myr.
        }}
        \label{fig:3_4}
    \end{figure}

\begin{figure}[ht!]
        \centering
        \includegraphics[
		width=\hsize]{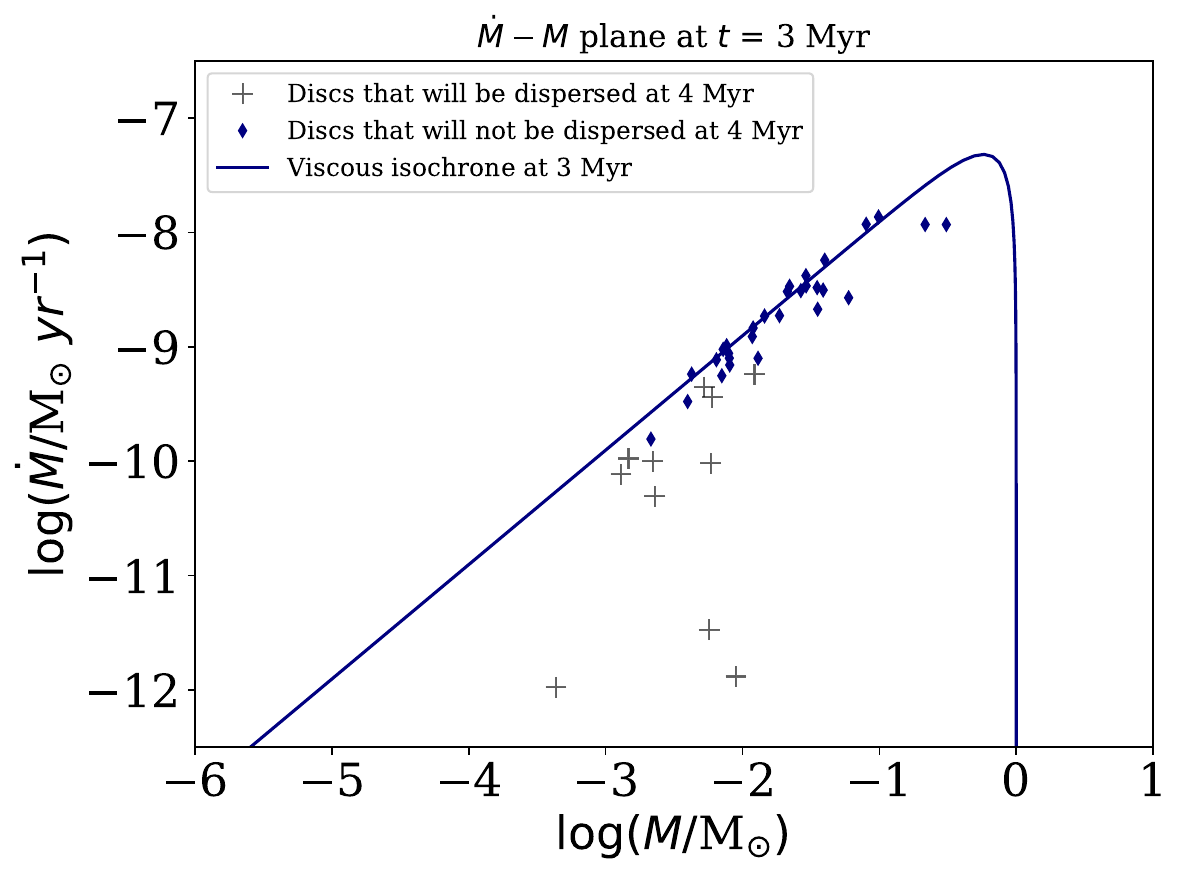}
        \caption{\small{Population in the $\dot{M}-M$ plane at 3 Myr compared with the viscous isochrone. Blue diamonds represent discs that are not dispersed at 4 Myr, while grey crosses discs that are dispersed at 4 Myr, which correspond to the shaded histogram in Figure \ref{fig:3_4}.
        }}
        \label{fig:isochrones}
    \end{figure}

\begin{figure}[ht!]
        \centering
        \includegraphics[
		width=\hsize]{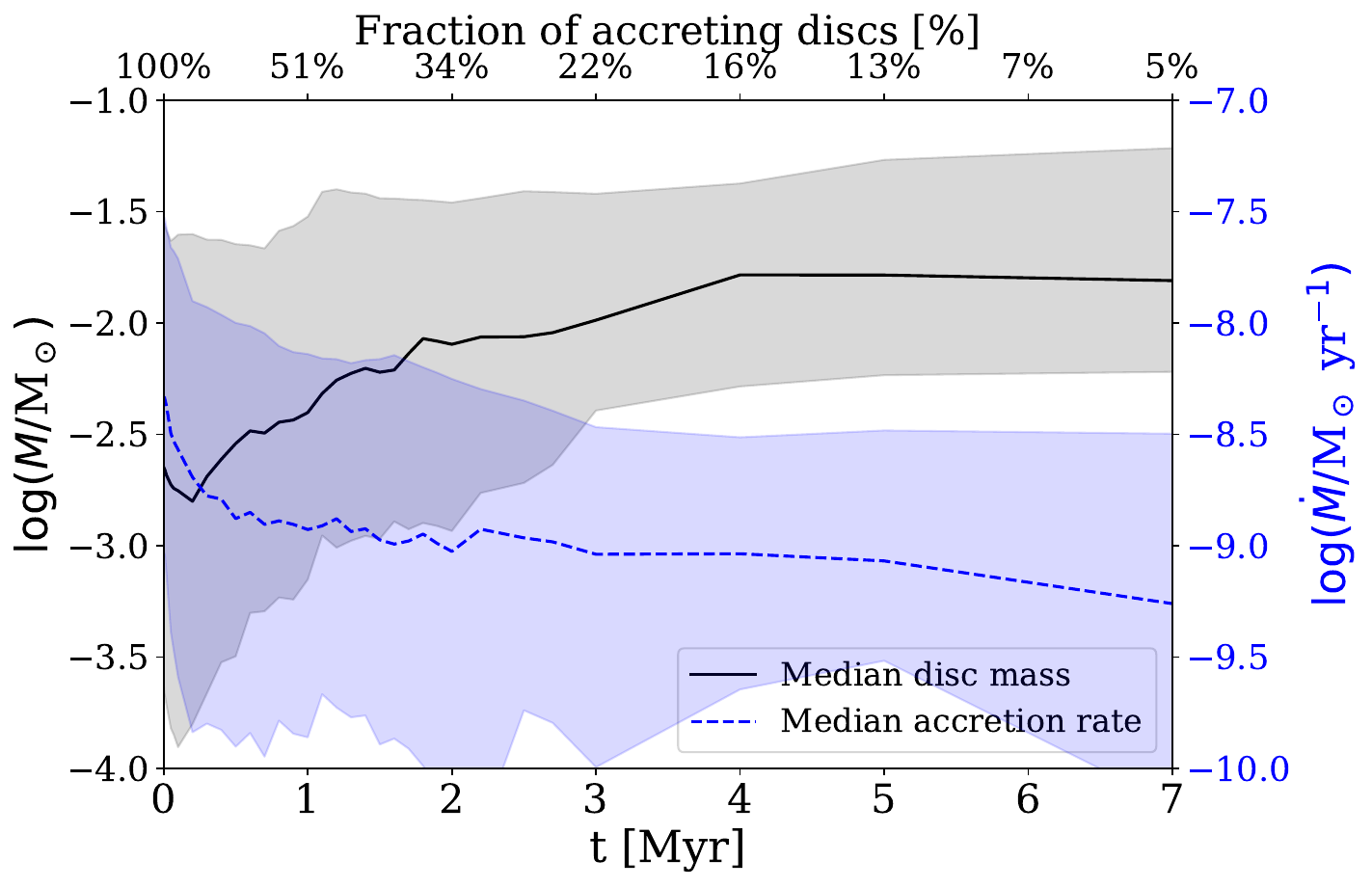}
        \caption{\small{Evolution of the median disc mass (solid black line) and the median accretion rate (dashed blue line) over time with shaded regions representing the one-sigma confidence intervals \rev{of the population.} The fraction of accreting discs at each timestep is shown above the plot.
        }}
        \label{fig:median_mass_median_accrate}
    \end{figure}

We expect to obtain the steepest increase of the median mass between 2 and 4 Myr, the interval during which most of the discs are dispersed.
We obtain that the generated population still manifests the increase of the median mass with time. Figure \ref{fig:3_4} illustrates the mass distribution of the population at 3 and 4 Myr, highlighting the increasing median and the fact that the discs dispersed between these two time snapshots, which are highlighted in grey, are the least massive in the population.
Hence, the removal of these low mass discs causes the older population to appear more massive than the younger one.
A further confirmation of the important role played by internal photoevaporation in this scenario is brought to light in Figure \ref{fig:isochrones}, which depicts the population in the $\dot{M}-M$ plane. The disc distribution in the plane is compared to the viscous isochrone \citep{Lodato2017}, which is the locus of points in the plane that the discs would occupy if their evolution were driven purely by viscous accretion.
Discs that are dispersed between 3 and 4 Myr (grey crosses) are placed below the curve, highlighting that their $\dot{M}$ is decreasing faster than $M$, while the purely viscous theory would predict a linear correlation between the two.
This feature is a hallmark of internal photoevaporation, predicted by \cite{Jones2012} and \cite{Rosotti2017} and found by \cite{Somigliana2020} in numerical simulations. Consequently, the fact that synthetic discs that show this behaviour are dispersed in the subsequent time snapshot highlights that internal photoevaporation is responsible for their depletion and for the increase of the median mass.
The fact that the accretion rate decreases more steeply than the mass, due to both accretion and internal photoevaporation, causes the median of the accretion rate distribution to decline over time.
This is highlighted in Figure \ref{fig:median_mass_median_accrate}, which shows the evolution of the median disc mass and the median accretion rate over time of the synthetic population studied in this section.
The decline of the accretion rate with the age of the population is in agreement with observations \citep{Testi2022}.
This survivorship bias, thus, is a specific property of the disc mass distribution.

\section{Discussion}
\label{discussion}

\subsection{The increase with time of the median mass - is it a general feature of viscous-photoevaporative models?}
\label{discussion_lambdas}

In Section \ref{model_correlations} \rev{we illustrated how crucial the impact of the slopes $\lambda_i$ of the correlations between the disc initial conditions and $M_\star$ is}, pointing out that if $0<\lambda<2$ (where $\lambda$ is defined in equation \ref{eqn:t_disp_corr1}) the median disc mass increases with time.
However, although some constraints of $\lambda_i$ exist in the literature \citep{Picogna2021, Somigliana2022}, they are not robust enough to rule out the possibility that $\lambda<0$ or $\lambda>2$, a scenario in which the increase of the median mass does not occur. In particular, the total mass-loss rate $\dot{M}_\text{w}$ has been obtained for a restricted sample of discs (see \citealt{Pascucci2023} for a recent review), which results in $\lambda_\text{w}$ being observationally unconstrained.
To properly evaluate which values of $\lambda_i$ most accurately reproduce the observed populations, one would need to conduct a population synthesis model with a robust statistical method to compare simulations with data.
Here we explore the impact of $\lambda_\text{w}$ in determining whether the increase of $M$ is a distinctive signature of this model.

The positive correlation between the X-Ray luminosity and the stellar mass \citep{Preibisch2005, Gudel2007, Ercolano2014} hints that negative $\lambda_\text{w}$ are not physically meaningful for low mass stars ($< 1.5 \ $M$_\odot$) for X-ray models.
If we restrict to the \cite{Somigliana2022} and \cite{Somigliana2024} constraints on $\lambda_{\mathrm{m},0}$ and $\lambda_{\text{acc},0}$ (see Section \ref{model_correlations}), and $\lambda_\text{w}>0$, we find that no combination allows $\lambda>2$ for either dispersal criteria.
However, it is possible to obtain $\lambda<0$, a condition under which massive discs are the first to be dispersed. 
We focus on the global criterion to determine the range of $\lambda_\text{w}$ for which the condition \rev{$0<\lambda<2$} holds because we find that, if this is the case, the condition holds also for the local criterion.
By assuming $\lambda_{\mathrm{m},0} = 2.1$ and $\lambda_{\text{acc},0} = 1.5$, which are the upper boundaries of the assumed constraints, we obtain that $\lambda<0$ only when $\lambda_\text{w}>2.4$, a rather steep $\dot{M}_\text{w}-M_\star$ correlation, not currently predicted by numerical simulations \citep{Picogna2021}.
Nevertheless, if we consider $\lambda_{\mathrm{m},0} = 1.2$ and $\lambda_{\text{acc},0} = 1.2$, \rev{which is the combination within the \cite{Somigliana2022} constraints for which $\lambda$ is lower,} we obtain a limit value of $\lambda_\text{w} = 1.2$, which is close to the $\rev{\lambda_\text{w}} \sim 1$ predicted by \cite{Picogna2021}. This suggests that a scenario where the median mass decreases with time cannot be completely excluded.
We can obtain additional constraints by considering the $R_\text{obs} - M_\star$ correlation, where $R_\text{obs}$ is the disc radius observed in the mm flux, which has been somewhat overlooked in the literature due to its lower prominence compared to the other two correlations and because of the significant uncertainties associated with $R_\text{obs}$ measurements \citep{Ansdell2018, Rosotti2019}. \cite{Andrews2018b}  and \cite{Hendler2020} found that $R_\text{obs} \propto {M_\star}^{0.6}$, which implies that massive stars tend to host larger discs. 
Furthermore, we know that the initial disc radius follows the relation
\begin{equation}
    \left(\frac{R_0}{\text{au}}\right) = 6 \pi \alpha \left(\frac{t_\nu}{\text{yr}}\right) \left(\frac{h}{r}\right)^2 \left(\frac{M_\star}{M_\odot}\right)^{-1/2} \ 
\end{equation}
from the viscous model, where $h/r = H/R$ at $R = 1 $ au and at $M_\star = 1$ M$_\odot$ and we assume that $\alpha$ does not depend on the stellar mass.
As a consequence, if we assume that the observed $R_\text{obs} - M_\star$ correlation is due to the $R_0$ distribution, we obtain 
\begin{equation}
    \lambda_{\mathrm{m},0} - \lambda_{\text{acc},0} \sim 1 \ .
\end{equation}
With this information, we conclude that $\lambda_\text{w}<2.2$ yields $\lambda>0$, which corresponds to an increase of the median mass of the population.
This value reflects that only steep $\dot{M}_\text{w}-M_\star$ correlations with $\lambda_\text{w}>2$ can prevent light discs to be the first to be dispersed in populations. 
As discussed in this section, numerical simulations currently predict flatter scalings of the mass-loss rate with $M_\star$ and, therefore, such steep correlations are very unlikely.
Thus, we can argue that, if viscosity drives accretion in discs and internal photoevaporation is responsible for their dispersal, older populations should appear more massive than younger ones, due to the survivorship bias.
However, we stress that this argument is based on the assumptions that the observed flux radius traces the disc critical radius $R_\text{c}$, which is non-trivial (see \citealt{Rosotti2019} and \citealt{Toci2021}).

\subsection{Outside-in dispersal}

\begin{figure}[ht!]
        \centering
        \includegraphics[
		width=\hsize]{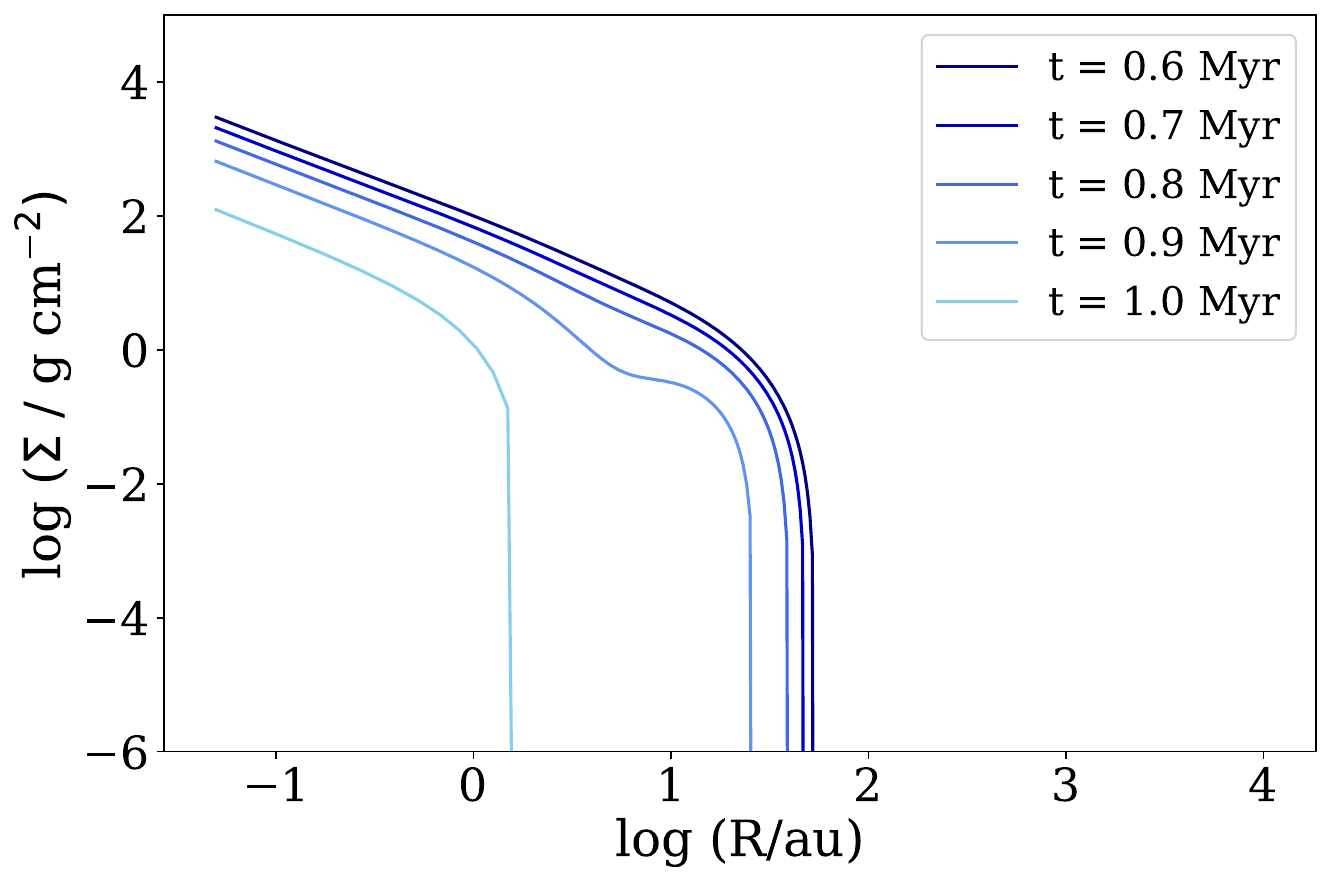}
        \caption{\small{Evolution of $\Sigma$ with time as a function of radius of a disc with $M_0 = 4 \cdot 10^{-3}$ M$_\odot$, $R_0 = 9.3$ au, $\alpha = 10^{-3}$ and $\dot{M}_\text{w} = 2.5 \cdot 10^{-9}$ M$_\odot$ yr$^{-1}$. For this specific disc, internal photoevaporation is more effective in the outer regions, resulting in an outside-in dispersal.
        }}
        \label{fig:outside_in}
    \end{figure}

\begin{figure}[ht!]
    \centering
    \includegraphics[
    width=\hsize]{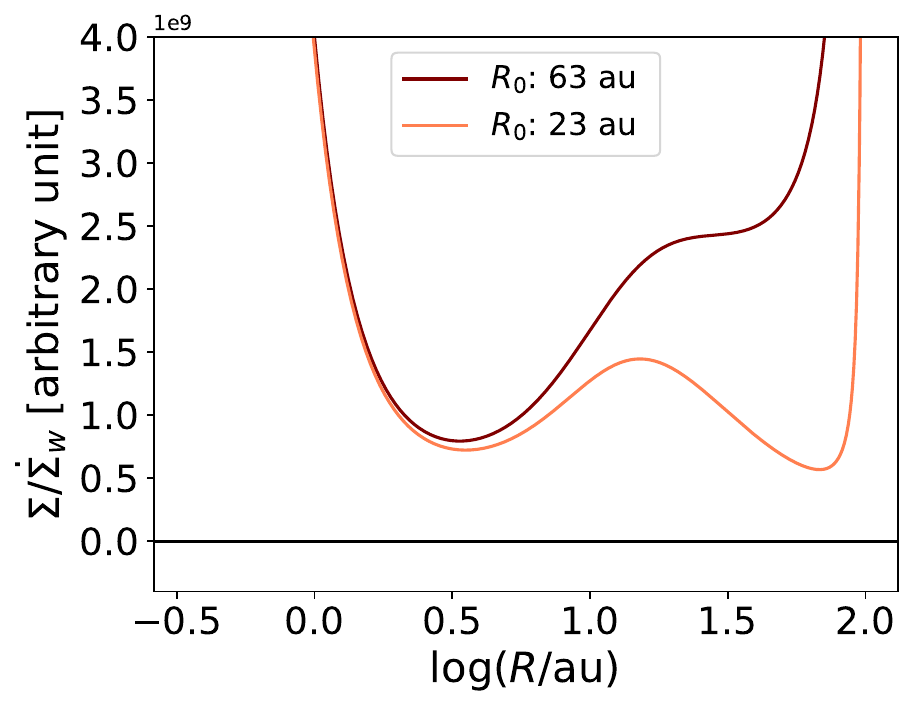}
    \caption{\small{Local dispersal timescale as a function of radius for a disc with $R_0 = 63$ au and a disc with $R_0 = 23$ au around a 1 M$_\odot$ star.}}
    \label{fig:sigmadotw_oi}
\end{figure}

The model described so far takes into account viscous accretion and internal photoevaporation, where we assumed throughout the paper that dispersal proceeds from inside out. 
This assumption has been crucial to state that the accretion rate was more affected than the disc mass by photoevaporation (see Figure \ref{fig:mass_accrate}) and, consequently, that the disc mass evolution could be described by purely viscous relations until the moment of disc dispersal.
This assumption allowed us to obtain an analytical relation \eqref{eqn:mass_distrib_vpe} that accurately describes the evolution of the $M$ distribution.
However, there are cases when internal photoevaporation can trigger an outside-in dispersal.
This feature is usually associated with external photoevaporation \citep{Clarke2007, Haworth2019, Winter2022, Anania2024} but can occur even for internal photoevaporation (\citealt{Ronco2024}, Tabone et al. in prep.).
Figure \ref{fig:outside_in} shows the evolution of the surface density of a disc affected by outside-in dispersal, triggered by internal photoevaporation.
In this particular case, with $R_0 \sim 10$ au, winds induce a mass-loss at large radii and deplete the outer region first. This results in a decreasing disc size and in the clearing of the disc from outside in.
To explain the physical reason behind this feature, we need to evaluate the dependence of the local dispersal timescale, defined in equation \eqref{eqn:t_loc_sigma}, on $R$. 
We assumed that the wind's mass-loss rate profile does not depend on the disc structure and always peaks at small radii (see equation \ref{eqn:peak_sigmadot}).
As a consequence, the $R$ dependence of $t_\text{loc}$ is introduced by the surface density profile \eqref{eqn:self_similar}.
When the truncation radius is considerably high, $t_\text{loc}$ has a minimum at $R_\text{gap}$, where $R_\text{gap}$ is defined in equation \eqref{eqn:peak_sigmadot}. 
The location of this minimum does not depend on $R_0$ but only on $M_\star$.
In compact discs the surface density is truncated at small radii, where the local mass-loss rate is still significant, which induces a second local minimum at large radii in the profile of $t_\text{loc}(R)$ (see Figure \ref{fig:sigmadotw_oi}).
The disc undergoes an outside-in dispersal when the second local minimum becomes a global minimum.
The only parameter that plays a key role in this is the truncation radius, which appears in the exponential term of equation \eqref{eqn:self_similar}.
We can obtain from numerical simulations the minimum radius $R'$ for which $t_\text{loc}$ has a global minimum at $R = R_\text{gap}$.
We find $R' \sim 26$ au by taking into account the \cite{Owen2012} $\dot{\Sigma}_\text{w}$ profile.
Less steep mass-loss profiles, such as the one obtained by 
\cite{Picogna2019}, where a larger fraction of mass is removed at large radii, produce a higher limit radius $R'$.
Consequently, discs with $R_\text{c}< R' \left(\frac{M_\star}{\text{M}_\odot}\right)$  at $t = t_\text{disp}$ do not open a gap but are depleted from outside-in.

Determining whether disc depletion occurs via inside-out (I-O) or outside-in (O-I) dispersal is not only crucial to assert whether the mass distribution increases with time - a feature predicted assuming I-O dispersal - but also to understand if winds are prone to carve cavities in gaseous discs.
Here we derive an analytical relation that separates the two regimes.
Let us consider the relation
\begin{equation}
    R_\text{c}(t_\text{disp}) < R' \left(\frac{M_\star}{\text{M}_\odot}\right) \ ,
    \label{eqn:io_oi}
\end{equation}
condition under which we expect O-I dispersal according to numerical simulations.
We consider the global criterion (see equation \ref{eqn:tw}) to evaluate $t_\text{disp}$, because the local criterion was obtained under the assumption of I-O dispersal, which is not valid in this scenario.
We account for viscous spreading by substituting
\begin{equation}
    R_\text{c}(t_\text{w}) = R_0 \left(1 + \frac{t_\text{w}}{t_\nu}\right)
    \label{eqn:viscous_spreading}
\end{equation}
if $\dot{M}_0 > \dot{M}_\text{w}$ and $R_\text{c} = R_0$ otherwise.
We substitute equation \eqref{eqn:viscous_spreading} in equation \eqref{eqn:io_oi} \rev{and} obtain the condition under which photoevaporation triggers an outside-in dispersal in evolved discs:
\begin{equation}
    M_0 < 2 \dot{M}_\text{w} t_\nu \left(\frac{R'}{R_0}\right)^{3/2}\left(\frac{M_\star}{\text{M}_\odot}\right)^{3/2} \ .
    \label{eqn:io_oi_}
\end{equation}
By substituting the typical values of disc parameters the condition becomes
\begin{align}
    M_0 < & \ 2.83 \cdot 10^{-3}  \left(\frac{\dot{M}_\text{w}}{10^{-9} \  \text{M}_\odot \ \text{yr}^{-1}}\right) \left(\frac{\alpha}{10^{-3}}\right)^{-1} \left(\frac{h/r}{1/30}\right)^{-2} \times \notag \\
    \times & \left(\frac{M_\star}{\text{M}_\odot}\right)^2 \left(\frac{R_0}{20 \ \text{au}}\right)^{-1/2} \left(\frac{R'}{26 \ \text{au}}\right)^{3/2} \ \text{M}_\odot \ .
    \label{eqn:io_oi_num}
\end{align}
This relation highlights that the more compact and the lighter the disc, the more likely it is to have outside-in dispersal. 
Moreover, the dispersal pathway strongly depends on the stellar mass and the shape of the $\dot{\Sigma}_\text{w}$ profile.
Equation \eqref{eqn:io_oi_num} is consistent with the results of Tabone et al. in prep., who obtained, with \texttt{Diskpop}, that small discs ($R_0 \sim 10$ au) undergo an outside-in dispersal.  
It is also consistent with \cite{Ronco2024}, who obtained numerically that O-I dispersal is typical of discs around massive stars, and with \cite{Coleman2022} who did not obtain this feature for discs with $R> 100$ au, unless affected by external photoevaporation.

\begin{figure}[ht!]
        \centering
        \includegraphics[
		width=\hsize]{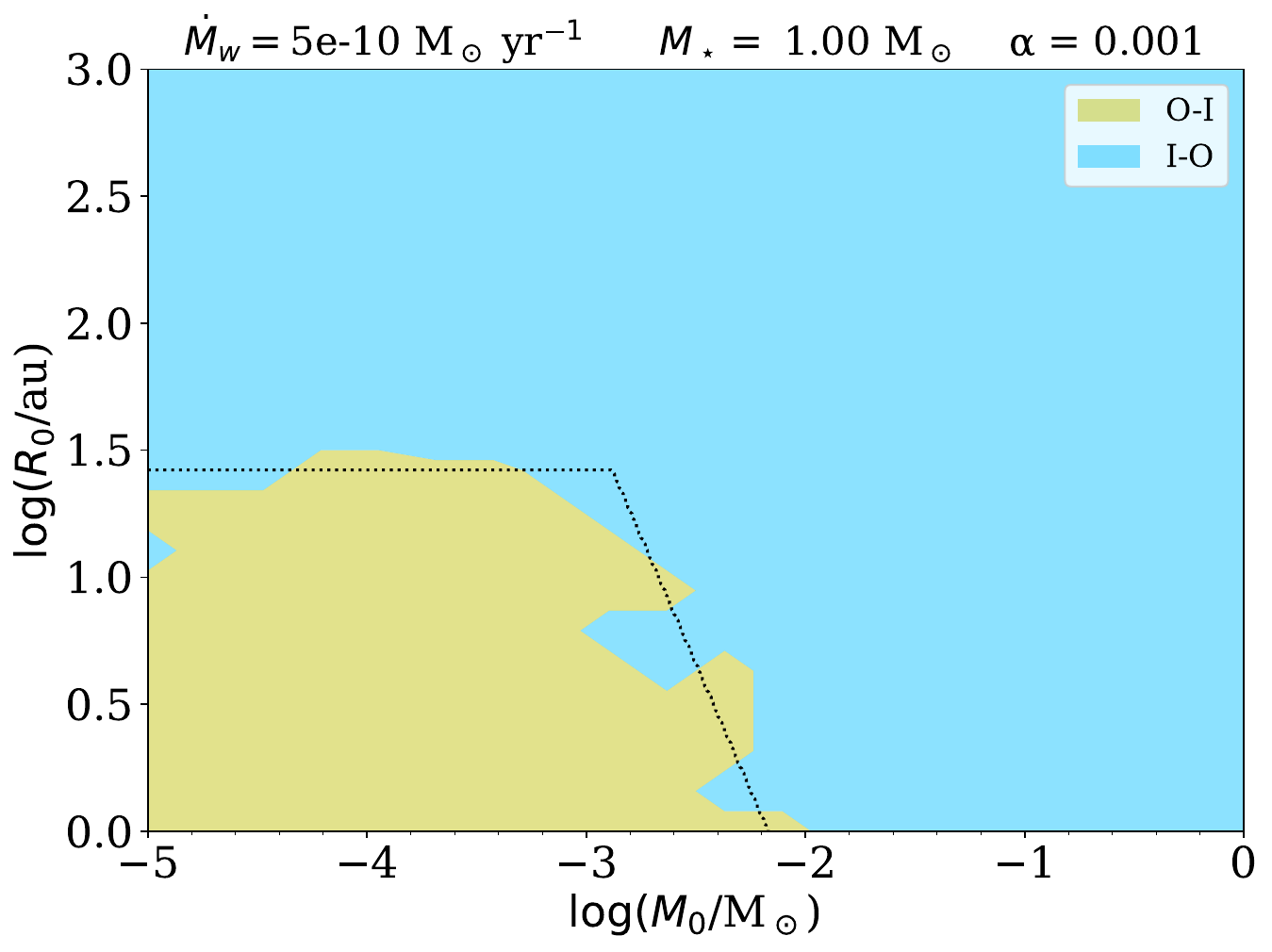}
        \caption{\small{Numerical test of equation \eqref{eqn:io_oi_num}. The colours represent the regions of the $R_0 - M_0$ plane where numerical discs are dispersed from outside in (in yellow) or from inside out (in blue). The dotted line represents the analytical boundary between these two regimes.
        }}
        \label{fig:check_outside_in}
    \end{figure}

\begin{figure}[ht!]
        \centering
        \includegraphics[
		width=\hsize]{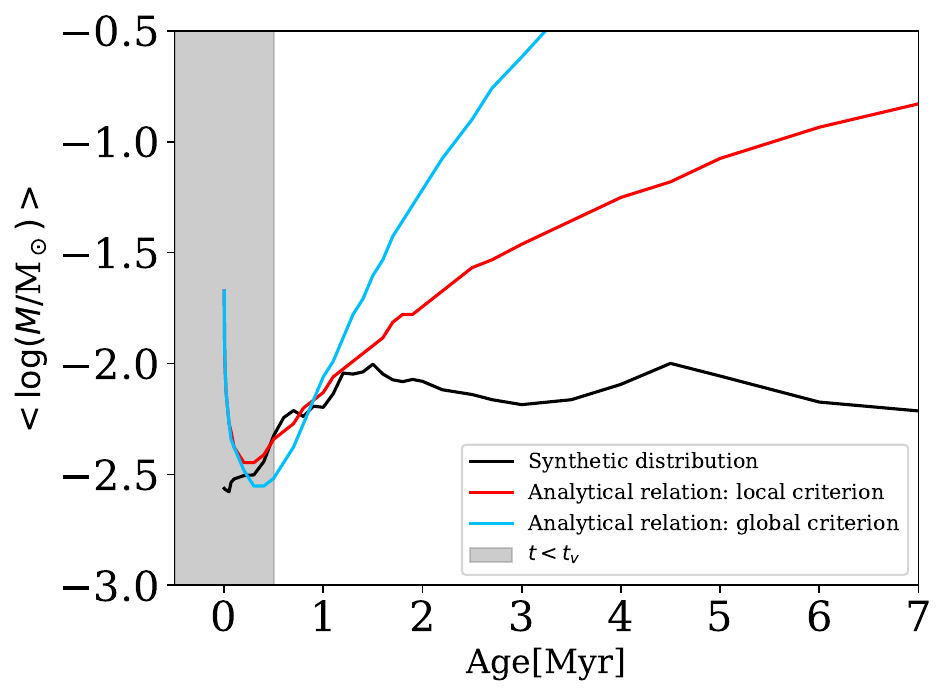}
        \caption{\small{Comparison between the evolution of the median mass with time of the numerically evolved population to the analytical predictions for both dispersal criteria. The shaded region corresponds to $t<t_\nu$, which is not considered in the model.
        }}
        \label{fig:medians_oi}
    \end{figure}

\begin{figure}[ht!]
        \centering
        \includegraphics[
		width=\hsize]{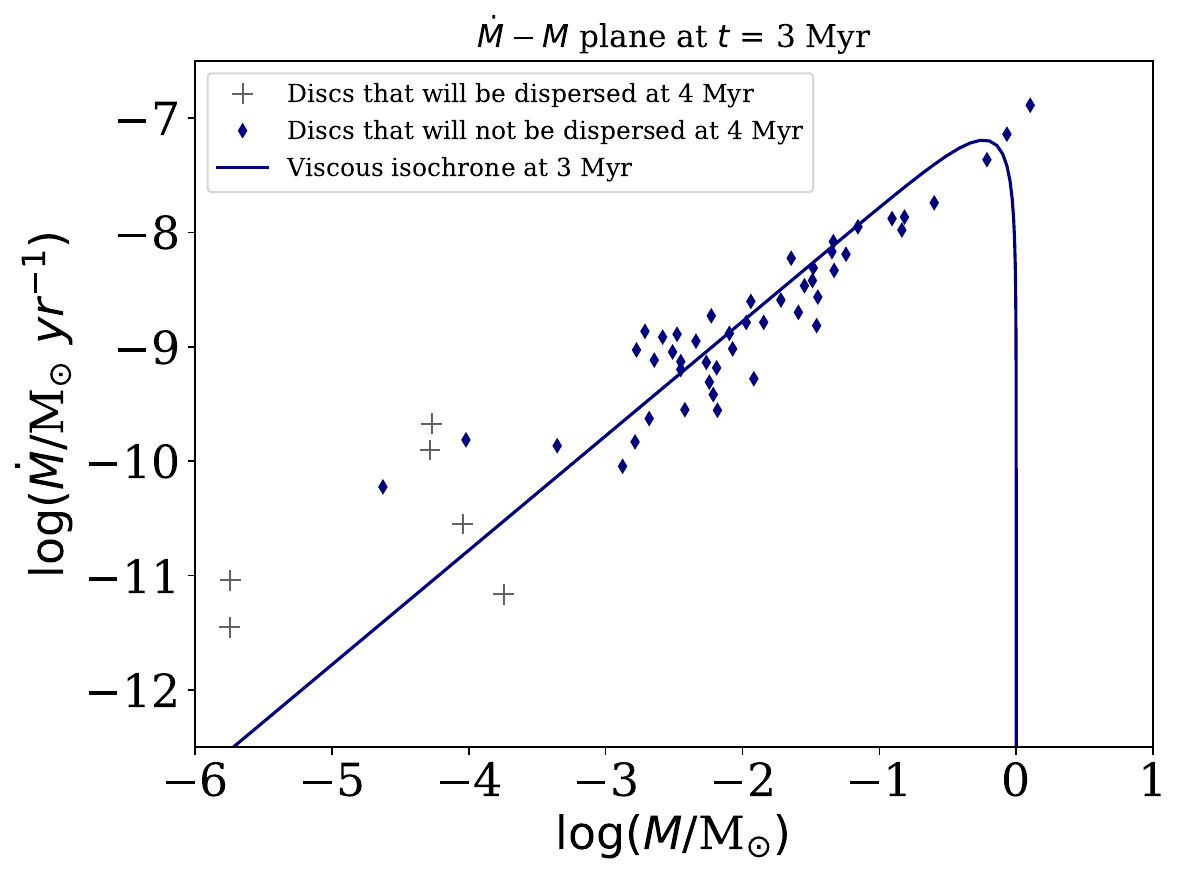}
        \caption{\small{
        Population in the $\dot{M}-M$ plane at 3 Myr compared with the viscous isochrone. Blue diamonds represent discs that are not dispersed at 4 Myr, while grey crosses discs that are dispersed at 4 Myr.
        }}
        \label{fig:isochrones_oi}
    \end{figure}

To evaluate the validity of equation \eqref{eqn:io_oi}, we performed different numerical simulations, varying only $M_0$ and $R_0$. We show in Figure \ref{fig:check_outside_in} that the analytical relation predicts the dispersal pathway with remarkable accuracy.
Then, we used \texttt{Diskpop} to analyse the mass evolution of a population where the majority of discs undergo outside-in disc dispersal.
This population has $\langle M_0 \rangle = 8 \cdot 10^{-3}$ M$_\odot$, $\langle \dot{M}_0 \rangle = 5 \cdot 10^{-9}$ M$_\odot$ yr$^{-1}$, $\langle \dot{M}_\text{w} \rangle = 3 \cdot 10^{-9}$ M$_\odot$ yr$^{-1}$ and $\alpha = 10^{-4}$.
We obtained that the mass distribution does not follow the analytical relation \eqref{eqn:mass_distrib_vpe} (see Figure \ref{fig:medians_oi}).
Such disagreement is expected, because O-I dispersal removes the outer regions first and, therefore, decreases the disc mass faster than the accretion rate.
Indeed, for these discs, the mass reaches its threshold before the accretion rate, in contrast with the result obtained with I-O dispersed discs (see Section \ref{realistic pop}).
As a consequence, not only O-I dispersal is slower than the I-O one, but also induces a mass decrease in the population, which balances the increase due to the removal of light discs.
Thus, the median mass stays somewhat constant between 2 and 4 Myr.
Figure \ref{fig:isochrones_oi} shows how this the analysed population is distributed in the $\dot{M} - M$ plane. It highlights that discs nearing depletion have their mass close to the threshold and are generally above the viscous isochrone, a signature of O-I dispersal \citep{Rosotti2017}.
In conclusion, whether photoevaporative dispersal occurs from inside out or outside in has a strong impact on the evolution of the disc mass.
It is indeed essential to understand which dispersal pathway is affecting observed disc populations.

\subsection{Comparison with the MHD winds model}
\label{mhd_model}

\begin{figure*}[ht!]
    \centering
    \includegraphics[width=0.33\linewidth]{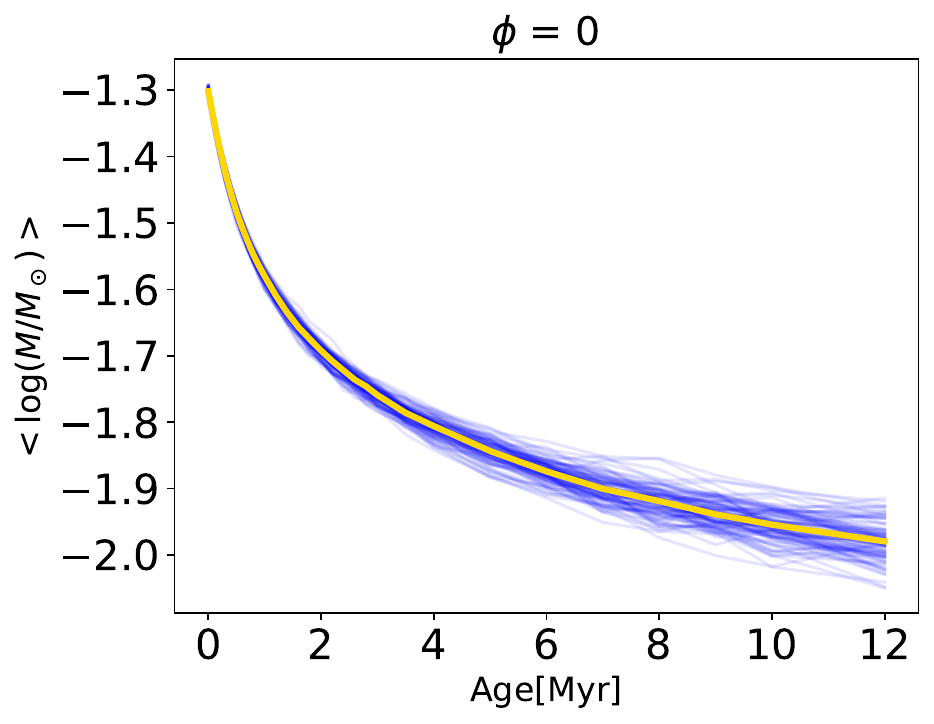}
    \includegraphics[width=0.33\linewidth]{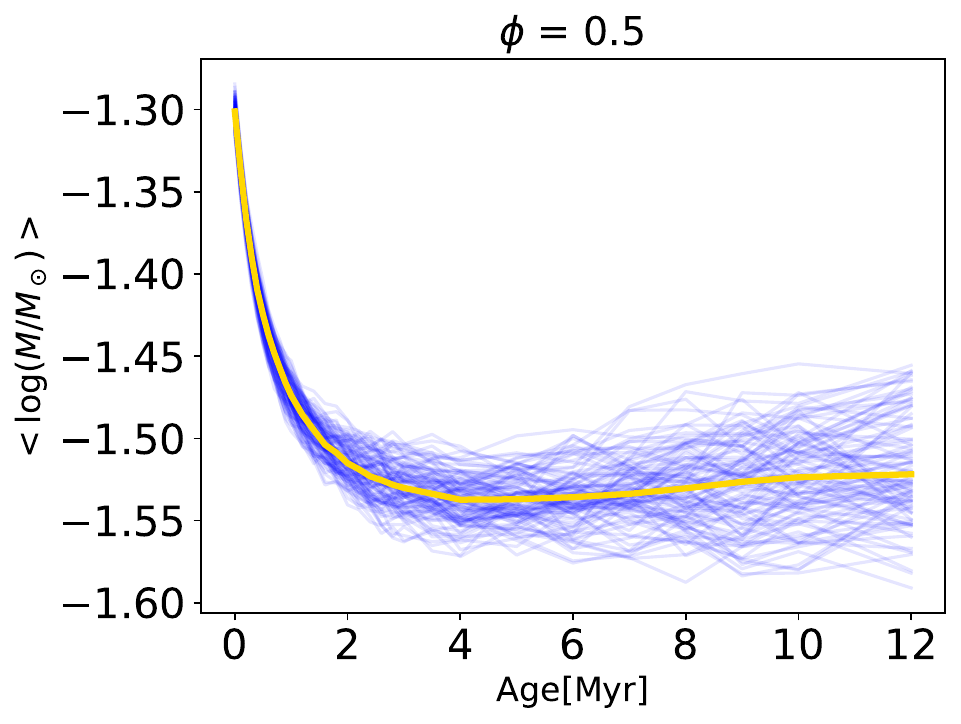} 
    \includegraphics[width=0.33\linewidth]{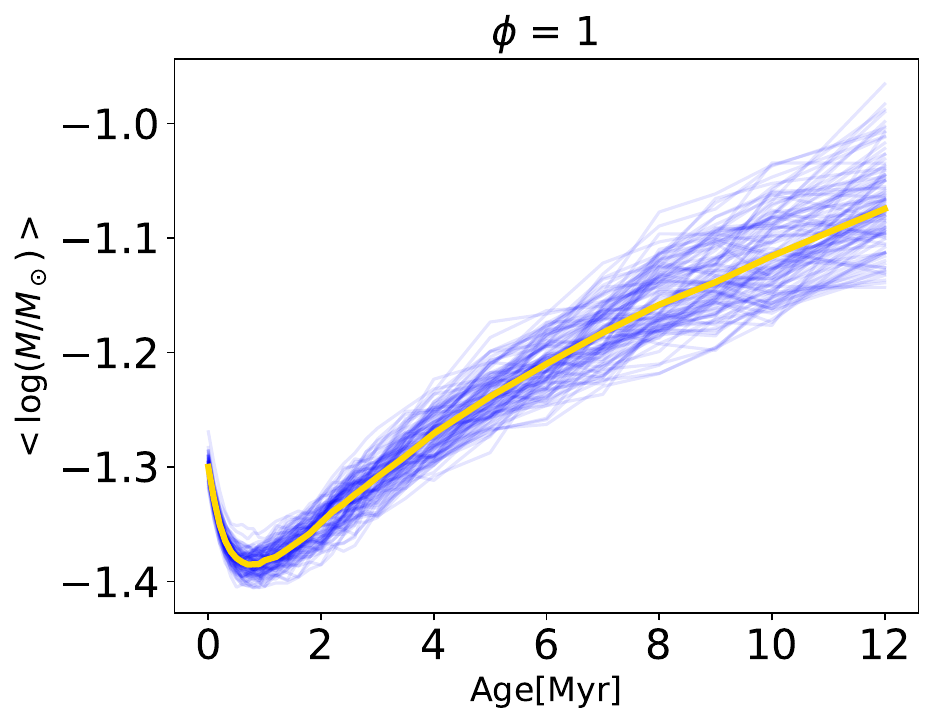} 
    \caption{Evolution of the median mass as a function of time for two sets of populations evolved according to the MHD winds model with $\phi = 0$ (left panel), $\phi = 0.5$ (central panel) and $\phi = 1$ (right panel).
    The golden lines represent the medians of all simulations with the same $\phi$.}
    \label{fig:mhd_correlations}
\end{figure*}

Unlike the viscous framework, the MHD winds model explicitly defines the dispersal time as the moment when both $M$ and $\dot{M}$ vanish \citep{Tabone2022model}.
This timescale depends on $t_\text{acc,0}$, a parameter that quantifies the accretion timescale.
A priori, this parameter is not correlated with $M_0$, which means that this model does not inherently predict that massive discs have a longer lifetime.
Consequently, \cite{Tabone2022} developed a population synthesis model where $M_0$ and $t_\text{acc,0}$ were extracted from independent distributions. They obtained that the median mass distribution decreases with the age of the population, because all discs lose mass due to accretion.
However, these results may change if there is an ad-hoc $M_0 - t_\text{acc,0}$ correlation
\begin{equation}
    M_0 \propto {t_\text{acc,0}}^\phi \ ,
\end{equation} 
with $\phi>0$, which implies that lighter discs have a shorter lifetime.
To explore this possibility, we simulated a population evolved with the MHD model, with the same model parameters as \cite{Zallio2024} and a correlation between $M_0$ and $t_\text{acc,0}$.
We find that only positive correlations with $\phi > 0.5$ cause the median mass to increase with time  (see Figure \ref{fig:mhd_correlations}).
However, $\phi>0.5$ is only marginally consistent with the results of \cite{Zallio2024} who found that, in order to reproduce the observed $\dot{M}-M$ correlation and the spread in $M$, MHD winds models must have $\phi \in [-2,1]$ (see Figure 12 of \citealt{Zallio2024}).
Therefore, we conclude that, although an increase is not completely ruled out, the region of the allowed parameter space for which MHD winds model predicts an increasing median disc mass is 
narrow.

\subsection{A new method to understand disc evolution from observed populations}
\label{Compare_observations}

Throughout this work, we showed that three different evolutionary pathways lead to three distinct signatures in the evolution of the median disc mass with time.
Indeed, the viscous framework generally predicts an increase of the median mass in case of inside-out (I-O) dispersal, and a constant median mass in case of outside-in (O-I) dispersal; whereas an MHD wind-driven accretion generally produces a decrease of such quantity over time.
This raises an important question: can observations of the disc mass shed light on the evolutionary mechanism that drives accretion?

The analytical model, described in Section \ref{analytic_m_distribution}, predicts an increase of the median mass as soon as the bulk of the discs are removed from the population, which corresponds to 2 - 3 Myr \citep{Hernandez2007, Fedele2010}. Therefore, our theoretical model predicts that this feature is   distinguishable between 2 and 4 - 6 Myr, where we expect an increase of the median mass by more than 0.5 dex (see Figure \ref{fig:medians}).
Since the median mass is less influenced by \rev{outliers} than the global mass distribution, this method is less affected by sample incompleteness compared to others (see e.g., \citealt{Alexander2023}, \citealt{Somigliana2024}).
\rev{As a result, we expect it to provide constraints also for samples of $\sim 100$ discs}. 
Consequently, a comparison between the median mass of ``young'' populations, such as Lupus and Chamaeleon, with $t \sim 2$ Myr \citep{Luhman2020, Galli2021}, and ``old'' populations, such as Upper-Sco, with 5 - 10 Myr \citep{Pecaut2016} is ideal for probing the accretion mechanism.
\cite{Testi2022} observed an increase of the continuum flux from 1 Myr to 2 Myr, followed by a decrease from 2 Myr to 5 Myr, which hints that the dust mass somewhat increases in the earliest phase of disc evolution. The authors interpreted this trend as a secondary dust production. 
Nevertheless, converting from continuum flux to gas mass remains non-trivial (see \citealt{Miotello2022} for a review), suggesting that gas measurements can probe more accurately the evolution of total disc mass.
Furthermore, different star-forming regions can be compared only if they can be described by the same initial conditions (see \citealt{Zagaria2023}).
To retrieve this information, we would need to implement a robust statistical method to compare synthetic populations with observed ones.
Moreover, as shown by \cite{Anania2024}, irradiation from massive stars produces non-negligible effects on the evolution disc sizes and masses in Upper-Sco, where both these quantities are reduced compared with the non-irradiated case. Therefore external photoevaporation must be included in the model to reproduce such star-forming region.
New large programs, such as AGE-PRO and DECO, will allow us to compare the expected evolution of the gas mass to the observed one.

\subsection{Caveats about $\dot{\Sigma}_\text{w}$ and $\alpha$}

In this section we discuss the assumptions that we have made throughout the paper about $\dot{\Sigma}_\text{w}$ and $\alpha$.
We assumed that the mass-loss rate does not depend on time and on the structure of the disc, following \cite{Owen2012}. However, \cite{Nakatani2021} found that the X-Ray luminosity of Pre-Main-Sequence stars can decrease by some order of magnitude as soon as the star enters the Main-Sequence. This effect is more pronounced for stars with $M_\star > 1$ M$_\odot$, while it remains negligible for T-Tauri stars, which we consider in this work.
Moreover, more recent mass-loss profiles \citep{Picogna2019, Nakatani2022, Sellek2024b} \rev{hint} that $\dot{\Sigma}_\text{w}$ \rev{should not be independent} on the disc parameters. In particular, the mass-loss rate undergoes a drop at $R>R_\text{c}$, which suggests that our analytical relation \eqref{eqn:io_oi_} \rev{might} overestimate the number of discs that are dispersed from outside in in populations.
\rev{However, to quantify the discrepancy between our model and more refined calculations, the dependence of $\dot{\Sigma}_\text{w}$ on $R_\text{c}$ should be investigated by numerical simulations.}
We foresee a future implementations of the updated $\dot{\Sigma}_\text{w}$ profiles in \texttt{Diskpop}.
In addition, although we separated MHD winds and photoevaporation in order to isolate their effects, these phenomena (as well as turbulence and MHD winds) are not mutually exclusive and can be present simultaneously in discs. Future works need to consider this assumption and study the effect of a hybrid model on the evolution of the disc mass distribution.

Throughout this work, we have always considered $\alpha$ as constant with respect to radius and stellar mass.
\cite{Garate2021} and \cite{Tong2024} explored the presence of a dead zone with small $\alpha$ at low radii and obtained a significant increase the lifetime of the inner disc. We expect such feature to slow down the decrease of $\dot{M}$ after the opening of the cavity and to lead to a slightly less steep increase of the median mass.
Various works (e.g., \citealt{Gorti2009b, Kunitomo2021}) introduced a $\alpha \propto M_\star$ scaling. This linear correlation affects the local criterion \eqref{eqn:t_loc}, which has a direct dependence on $\alpha$, but also the global criterion because $t_\nu = t_\nu(\alpha)$. 
As a consequence, if this correlation existed, the constraint given by $R_\text{obs}$ will become $\lambda_{\text{m},0} - \lambda_{\text{acc},0} \sim 0$, which reduces the parameter region in which the increase of the median mass is expected.

Finally, we stress that our work neglects planet-disc interactions, which become significant if protoplanets had already formed by $t \sim 2$ Myr.

\section{Conclusions}
\label{conclusions}

In this paper we explored the evolution of the mass distribution of a protoplanetary disc population that evolves according to the viscous-photoevaporative framework.
We derived an analytical relation that describes it and discussed its physical meaning.
To verify the analytical predictions, we performed various numerical simulations with \texttt{Diskpop} \citep{Somigliana2024} and confirmed our expectations.
Then, we compared our results with different evolutionary pathways such as outside-in \rev{photoevaporative} dispersal and MHD winds evolution.
Our main results are the following:
\begin{enumerate}
    \item Populations evolved assuming viscous accretion and undergoing an inside-out dispersal driven by internal photoevaporation exhibit a unique feature: the apparent increase with time of the median disc mass. 
    This occurs because lighter discs are the first to be dispersed by internal photoevaporation. This form of survivorship bias is manifested by populations where disc properties are independent of each other and, across a broad region of the parameter space, it also occurs in populations where each quantity is correlated with the stellar mass. The increase of the median disc mass does not imply a rise in the median mass accretion rate in populations; instead, the accretion rate declines over time.
    \item Other evolutionary pathways do not manifest the same evolution of the median disc mass. In particular, an outside-in dispersal results in the median mass remaining somewhat constant over time, while the MHD winds model in general predicts it to decrease with time.
    Such difference highlights that the increase of the median mass is a signature of the viscous-photoevaporative model with inside-out dispersal.
    \item We derived a new criterion to estimate the disc's lifetime as a function of initial disc properties. We further obtained an analytical relation that predicts that internal photoevaporation triggers an inside-out (outside-in) disc dispersal if its critical radius at the moment of dispersal is larger (smaller) than a numerically determined threshold. We verified both analytical relations with numerically evolved synthetic populations.
    \item We propose a new method to distinguish between the viscous-photoevaporative and MHD wind-driven evolution in observations. Indeed, comparing the mass distribution of 2 Myr old observed populations, such as Lupus, and older populations, such as Upper-Sco, could allow us to disentangle the accretion mechanism. For this scope, the recent large programs AGE-PRO and DECO, which aim to expand the sample of gas masses in disc populations, provide a valuable opportunity.
\end{enumerate}
Future studies should concentrate on providing a statistically robust comparison between observed populations and theoretical expectations. This will enable us to verify if young and old star-forming regions are comparable. Moreover, measuring disc masses with more accuracy and precision (see for example \citealt{Lodato2023}, \citealt{Martire2024}) will be crucial for understanding disc evolution.

\begin{acknowledgements}

The authors acknowledge support from the European Union (ERC Starting Grant DiscEvol, project number 101039651 and ERC, WANDA, project number 101039452), from Fondazione Cariplo, grant No. 2022-1217, from the European Union’s Horizon 2020 research and innovation programme under the Marie Sklodowska-Curie grant agreement No 823823 (Dustbusters RISE project), and from the ERC Synergy Grant “ECOGAL” (project ID 855130). 
Views and opinions expressed are, however, those of the authors only and do not necessarily reflect those of the European Union or the European Research Council. Neither the European Union nor the granting authority can be held responsible for them.
Part of this work was supported by the European Southern Observatory (ESO) who funded LM for a three-month internship.
GL acknowledges financial contribution from PRIN-MUR 20228JPA3A.
Moreover, we thank Cathie Clarke and Giovanni Picogna for the insightful discussions and very useful comments.

\end{acknowledgements}

\bibliographystyle{aa}

\bibliography{repbib}

\clearpage

\begin{appendix}

\renewcommand{\theequation}{A.\arabic{equation}}
\setcounter{equation}{0}

\section*{Appendix A: Derivation of the analytical mass distribution}
\label{appendix}   

Here we show how we derived the analytical mass distribution (see Figure \ref{fig:cutgaussian}).
Let us consider equation \eqref{eqn:mass_distrib_vpe}, where 
\begin{equation}
    \frac{\partial N}{\partial m_0} = N_1 \exp\left(-\frac{(m_0 - \overline{m_0})^2}{2 {\sigma_{m_0}}^2}\right) \ 
\end{equation}
and 
\begin{equation}
    \frac{\partial N}{\partial \dot{m}_0} = N_2 \exp\left(-\frac{(\dot{m}_0 - \overline{\dot{m}_0})^2}{2 {\sigma_{\dot{{m}_0}}^2}}\right) \ .
\end{equation}
The relation becomes
\begin{equation}
    \frac{\partial N}{\partial m} = N_1 N_2 \int^{+ \infty} _{m_{0,\text{MIN}}} \text{d} \tildea{m_0} \exp\left(-\frac{1}{2} \mathcal{M}\right) \ ,
\end{equation}
with 
\begin{equation}
    \mathcal{M} = \frac{\tilde{m_0}^2}{{\sigma_{m_0}}^2} + \frac{9}{{\sigma_{\dot{m}_0}}^2}\left(\tildea{m_0}- \frac{2}{3}\tildea{m}\right)^2 \ ,
\end{equation}
$\tildea{m_0} \equiv m_0 - \overline{m_0}$ and $\tildea{m} \equiv m - \overline{m}$.
Following \cite{Somigliana2022}, we isolate the dependence of the integration variable $\tildea{m_0}$, so that we can move the term that depends only on $\tildea{m}$ out of the integral;
\begin{equation}
    \frac{\partial N}{\partial m} = N_3 \exp \left(- \frac{\tildea{m}^2}{2 {\sigma}^2}\right) \cdot \int^{+\infty}_{\tildea{m}_{0,\text{MIN}}} \text{d} \tildea{m_0} \exp \left[-\frac{1}{2}\left(A \tildea{m_0}- \frac{2}{3}\frac{B^2} {A}\tildea{m}\right)\rev{^2}\right] \ ,
\end{equation}
where $\sigma^2 = \frac{9}{4} {\sigma_{m_0}}^{2} + \frac{1}{4} {\sigma_{\dot{m}_0}}^{2}$, $A^2 = \frac{1}{{\sigma_{m_0}}^{2}} + \frac{9}{{\sigma_{\dot{m}_0}}^{2}}$ and $B^2 = \frac{9}{{\sigma_{\dot{m}_0}}^2}$.
We notice that the exponential term outside the integral matches to the one obtained by \cite{Somigliana2022} considering the purely viscous model.
This term represents a Gaussian distribution whose mean is constantly decreasing with time due to accretion.
Instead, photoevaporation only introduces a minimum initial mass $m_{0,\text{MIN}}$, without which the second term would be a mere multiplicative constant.
After some straightforward integration we obtain
\begin{align}
    \frac{\partial N}{\partial m} = \ & N_3 \exp \left(- \frac{(m - \overline{m}(t))^2}{2 {\sigma}^2}\right) \times \notag \\
    \times & \sqrt{\frac{\pi}{2}} \ \frac{1 -\text{erf}\left[\frac{1}{\sqrt{2}}A(m_{0,\text{MIN}}(t)-m_0)- \frac{2}{3}\frac{B^2}{A}(m-\overline{m})\right]}{A} \ .
\end{align} 
\end{appendix}

\end{document}